\documentclass[twocolumn]{aastex631}

\submitjournal{ApJ}

\usepackage{outlines}
\usepackage{tablefootnote}
\usepackage{booktabs}        
\usepackage{amsmath}
\usepackage{makecell}

\usepackage{multirow}
\usepackage[utf8]{inputenc}
\usepackage[english]{babel}
\usepackage{pagecolor,lipsum}

\makeatletter
\newcommand{\globalcolor}[1]{%
  \color{#1}\global\let\default@color\current@color
}
 \newcommand\addressresponse[1]{#1}
\makeatother

\newcommand*{\linktocite}[2]{%
  \hyper@natlinkstart{#1}#2\hyper@natlinkend} 

\shorttitle{Diagnosing limb asymmetries in transmission}
\shortauthors{Savel et al.}

\graphicspath{{./}{figures/}}

\begin{document}

\title{Diagnosing limb asymmetries in hot and ultra-hot Jupiters with high-resolution transmission spectroscopy}

\newcommand\umd{\affiliation{Astronomy Department, University of Maryland, College Park, 4296 Stadium Dr., College Park, MD 207842 USA}}
\newcommand\cca{\affiliation{Center for Computational Astrophysics, Flatiron Institute, New York, NY 10010, USA}}
\newcommand\umich{\affiliation{Department of Astronomy, University of Michigan, 1085 South University Avenue, Ann Arbor, MI 48109, USA}}
\newcommand\uchicago{\affiliation{Department of Astronomy \& Astrophysics, University of Chicago, Chicago, IL 60637, USA}}
\newcommand\asu{\affiliation{School of Earth and Space Exploration, Arizona State University, PO Box 871404, Tempe, AZ 85281, USA}}

\correspondingauthor{Arjun B. Savel}
\email{asavel@umd.edu}

\author[0000-0002-0786-7307]{Arjun B. Savel}
\cca
\umd

\author[0000-0002-1337-9051]{Eliza M.-R. Kempton}
\umd

\author[0000-0003-3963-9672]{Emily Rauscher}
\umich

\author[0000-0002-9258-5311]{Thaddeus D. Komacek}
\umd

\author[0000-0003-4733-6532]{Jacob L. Bean}
\uchicago

\author[0000-0002-2110-6694]{Matej Malik}
\umd

\author[0000-0003-0217-3880]{Isaac Malsky}
\umich


\begin{abstract}
Due to their likely tidally synchronized nature, (ultra)hot Jupiter atmospheres should experience  strongly spatially heterogeneous instellation. The large irradiation contrast and resulting atmospheric circulation induce temperature and chemical gradients that can produce asymmetries across the eastern and western limbs of these atmospheres during transit. By observing an (ultra)hot Jupiter's transmission spectrum at high spectral resolution, these asymmetries can be recovered---namely through net Doppler shifts originating from the exoplanet's atmosphere yielded by cross-correlation analysis. Given the range of mechanisms at play, identifying the underlying cause of observed asymmetry is nontrivial. In this work, we explore sources and diagnostics of asymmetries in high-resolution cross-correlation spectroscopy of hot and ultra-hot Jupiters using both parameterized and self-consistent atmospheric models. If an asymmetry is observed, we find that it can be difficult to attribute it to equilibrium chemistry gradients because many other processes can produce asymmetries. Identifying a molecule that is chemically stable over the temperature range of a planetary atmosphere can help establish a ``baseline'' to disentangle the various potential causes of limb asymmetries observed in other species. We identify CO as an ideal molecule, given its stability over nearly the entirety of the ultra-hot Jupiter temperature range. Furthermore, we find that if limb asymmetry is due to morning terminator clouds, blueshifts for a number of species should decrease during transit. Finally, by comparing our forward models to \cite{kesseli2022atomic}, we demonstrate that binning high-resolution spectra into two phase bins provides a desirable trade-off between maintaining signal to noise and resolving asymmetries.
\end{abstract}



\section{Introduction} \label{sec:intro}

Exoplanet atmospheres vary spatially. This is especially the case for tidally locked exoplanets, which feature permanent daysides and permanent nightsides; such strong gradients in instellation in turn drive strong latitudinal and longitudinal variations in atmospheric temperature, dynamics, and chemistry \citep[e.g.,][]{showman2002atmospheric,cooper2005dynamic,harrington2006phase,cho2008atmospheric, menou2009atmospheric,showman2009atmospheric, rauscher2010three,perna2012effects,mayne2014unified,demory2016map,fromang2016shear,kataria2016atmospheric, parmentier2018thermal,zhang2018global, kreidberg2019absence,komacek2019vertical,  tan2019atmospheric,parmentier2021cloudy, roman2021clouds}. 


The spatial variations in exoplanet atmospheres have increasingly observable ramifications. Even with the insight gained from modeling exoplanet atmospheres as one-dimensional objects \citep[e.g.,][]{madhusudhan2009temperature,crossfield2017trends,yan2019ionized, benneke2019water}, a substantial and growing literature demonstrates that upcoming JWST \citep{beichman2014observations} data will require consideration of 3D processes for accurate interpretation of exoplanet atmospheric data \citep{feng2016impact, blecic2017implications,caldas2019effects, lacy2020jwst,macdonald2020so, pluriel2020strong, espinoza2021constraining, pluriel2021towards,macdonald2022trident,welbanks2022atmospheric}. Perhaps more urgently, ground-based high-resolution ($R \geq 15,000$) cross-correlation spectroscopy \citep[HRCCS; for a review, see][]{birkby2018exoplanet} datasets already show signs of significant multidimensionality \citep{flowers2019high, beltz2020significant, gandhi2022spatially, herman2022dayside, van2022carbon}. 

In transit geometry, HRCCS is similar to the more traditional transmission spectroscopy technique \citep[e.g.,][]{charbonneau2002detection}. Both methods leverage the idea that, as an exoplanet passes between its host star and an observer, stellar light is attenuated on a wavelength-dependent basis as it passes through the upper layers of the planet's atmosphere. But with HRCCS, the planetary absorption spectrum is buried in the stellar and telluric noise. Therefore, models of planetary absorption often cannot be directly compared to HRCCS data.\footnote{There are a few notable exceptions in the high-resolution spectroscopy literature in which planetary absorption is strong enough (and data quality high enough) that individual planetary absorption lines can be analyzed; e.g., \cite{tabernero2021espresso}.} However, by leveraging cross-correlation techniques, researchers can combine the signal from the many planetary absorption lines resolved at high resolution to yield a combined, statistically significant signal \citep[e.g., ][]{snellen2010orbital}.

The resolving of individual spectral lines allows for more than just binary detection/non-detection of planetary absorption: crucially, the Doppler shifts of planetary absorption lines are recoverable. The Doppler shifting of planetary lines due to the planet's orbital motion is in fact central for extracting the planetary signal with cross-correlation techniques, as the stellar and telluric lines are comparatively static. Specifically, a \textit{template spectrum} is chosen to model the planetary absorption signal, and it is cross-correlated against the combined planet, star, and telluric signal by Doppler shifting the template at varying velocities and multiplying the shifted template against the combined observed signal. The resulting cross-correlation function (CCF, a function of Doppler-shifted velocity) is maximized at the Doppler shift where the template best matches the combined observed signal---that is, at the Doppler shift of the planet signal in the observed combined data. Again, this method requires that the planet's spectral lines move across a spectrograph's pixels during observations, with the stellar and telluric lines largely remaining on the same pixel (or being easily detrended in time). With current instruments, this assumption is certainly justified for tidally locked ultra-hot Jupiters, which tend to have high orbital velocities \citep[e.g.,][]{fortney2021hot}.

With the planetary signal identified, further Doppler shifting and line broadening that is not associated with planetary orbital motion, telluric lines, or stellar lines is attributable to the 3D manifestations of planetary rotation and winds \citep{kempton2012constraining, showman2012doppler, kempton2014high, brogi2016rotation, ehrenreich2020nightside}. Thus, the multidimensionality of exoplanetary atmospheres is imprinted on HRCCS data.

Recent years have seen the intrinsic 3-dimensionality of these objects be uniquely constrained with transit HRCCS results. Observational studies such as \cite{louden2015spatially},  \cite{ehrenreich2020nightside}, and \cite{kesseli2022atomic} have isolated signals from the morning and evening limbs of planetary atmospheres, unveiling Doppler shifts of multiple chemical species at multiple points in transit---and hence over multiple longitudinal slices. Such studies have revealed asymmetries in the probed Doppler velocity field (i.e., changes in the Doppler shift of the CCF maximum as a function of orbital phase), which are often attributed to physical asymmetries in the atmosphere.

However, as reviewed in Section~\ref{sec:drivers}, an asymmetric signal in HRCCS can arise from a combination of different classes of mechanisms: 1) chemistry, 2) clouds, 3) dynamics, 4) orbital properties, and 5) thermal structure. Disentangling these effects is not a straightforward process. This may especially be the case if transmission spectra must be stacked together to achieve a higher signal-to-noise ratio (SNR), thereby smearing phase information.

In this work, we aim to explore the general question of asymmetry in exoplanet atmospheres, with particular focus on its manifestations in high-resolution transmission spectroscopy. Section~\ref{sec:drivers} examines what drives asymmetry in exoplanet atmospheres; we here define a metric that quantifies limb-to-limb asymmetry. In Section~\ref{sec:diagnostics}, we elaborate on diagnostics of specific mechanisms that may drive such asymmetries. This section additionally emphasizes how these diagnostics may be used to support or falsify compelling ``toy models'' motivated by the drivers described in Section~\ref{sec:drivers}. Finally, we summarize our results in Section \ref{sec:conclusion}.

\begin{table*}[h]
\centering
\caption{Example drivers of phase asymmetries of ultra-hot Jupiters}
\label{table:tests}
\centering
\begin{tabular}{c|c|c}
\toprule
 \textbf{Mechanism} &
\textbf{Proposed diagnostic} & \textbf{Selected work(s)}  \\
\bottomrule
Escaping atmosphere & H Lyman-$\alpha$ transit duration& \cite{owen2023fundamentals} \\
 & Very deep, e.g., $\rm Ca\, II$ lines & \cite{fossati2013absorbing}\\
 & Strong vertical winds & \cite{seidel2021into}\\
 & Very broadened, e.g., Na I lines & \cite{hoeijmakers2020hot}\\
 & Large diff. between (Na) doublet lines & \cite{hoeijmakers2020hot} \\
 & Strong H-$\alpha$ absorption & \cite{wyttenbach2020mass} \\
 & Blueshifting CCF w/ $\rm phase^P$ & \cite{bourrier2020hot} \\
 & Excess He absorption (10830\AA)& \cite{oklopvcic2018new}, \cite{spake2018helium}\\
 & Compare Doppler shifts of ions w/ different masses& This work\\
\hline
Scale height difference & Blueshifting CCF w/ phase & \cite{kempton2012constraining}\\
 & Strongly varying CO Doppler shift w/ phase & \cite{wardenier2021decomposing}, this work \\\hline

$\rm H_2$ dissoc./recomb. & Small phase curve amplitude & \cite{mansfield2020evidence}\\
 & High continuum/muted spectrum from $\rm H^-$& \cite{arcangeli2018h}\\
\hline

Weak drag state & Blueshifted CCF & \cite{wardenier2021decomposing}, \cite{savel2022no} \\
 & Large phase curve offset & {May \& Komacek et~al.} (\citeyear{maykomacek2021spitzer}) \\\hline
Cold interior & Blueshifted CCF & \cite{savel2022no} \\
 & Large phase curve amplitude & {May \& Komacek et~al.} (\citeyear{maykomacek2021spitzer}) \\
 & Cold nightside & {May \& Komacek et~al.} (\citeyear{maykomacek2021spitzer})\\\hline
Superrotating jet & CCF FWHM exceeding solid-body rotation & \cite{brogi2016rotation} \\
 & Phase curve offset & \cite{knutson2007map}\\\hline
Day-night winds & Blueshifted CCF & \cite{snellen2010orbital}\\
\hline
Equilibrium chemistry & Limb-to-limb abundance discrepancy & This work \\
 & Compare chemical, dynamical timescales & \cite{showman2012doppler}\\\hline
Photochemistry & Disequilibrium abundance of species & \cite{tsai2017vulcan} \\
 & Increase in product, decrease in parent w/ $\rm phase^P$ & Future work \\\hline
Condensation  & Model with GCM & \cite{wardenier2021decomposing}, \cite{savel2022no}\\
 & Strongly blueshifting CCF w/ phase & \cite{ehrenreich2020nightside} \\\hline
Eccentricity & All lines similarly Doppler shifted & \cite{montalto2011exoplanet}, \cite{savel2022no}\\
 & Independent orbital constraints & \cite{montalto2011exoplanet}\\\hline
Clouds & All species become less blueshifted w/ phase  & This work \\
 & Blueshifted CCF  & \cite{savel2022no} \\
 & Lines absent in low resolution present at high resolution & \cite{kempton2014high}, \cite{hood2020prospects}\\
 & Comparing CCFs of water bands & \cite{pino2018diagnosing}\\\hline
Lorentz forces & Reduced hotspot offset & \cite{beltz2021exploring} \\
 & Increased phase curve amplitude & \cite{beltz2021exploring} \\
 & Westward hotspot offset & \cite{hindle2021magnetic} \\\hline
Spatially varying winds & Variable Doppler shift over phase & \cite{kempton2012constraining}\\
 & Compare ingress / egress Doppler shifts& \cite{kempton2012constraining} \\
 & Compare Doppler shifts of different-strength lines& \cite{kempton2012constraining}\\\hline
Tidal deformation/lag & Blueshifting CCF over $\rm phase^P$ & Future work \\ 
& Light curve fitting & \cite{akinsanmi2019detectability} \\\hline
T-dependent opacity & Blueshifting CCF over phase & \cite{wardenier2021decomposing} \\ \hline
\end{tabular}
\raggedright
\tablecomments{Tests with a ``P'' superscript have been proposed but not explicitly modeled.}
\end{table*}

\section{Selected drivers} \label{sec:drivers}
There exist a number of potential drivers of asymmetry in high-resolution transmission spectroscopy. But what are the relative strengths of these drivers?

Previous works have considered the effects of condensation, longitude-dependent winds, and orbital eccentricity in producing such asymmetries \citep{wardenier2021decomposing, savel2022no}. Table~\ref{table:tests} includes these and a number of other potential drivers of asymmetry (along with potential diagnostics; Section~\ref{sec:diagnostics}). While many drivers are listed in Table~\ref{table:tests}, we consider in this work the relative strengths of two potentially first-order effects: the ``scale height effect'' and differences in equilibrium chemistry abundance across the limbs of the planet. Being both temperature-dependent effects, the distinction between the two is particularly subtle from an observational perspective, and hence interesting from a theoretical perspective.

The scale height effect is due to the larger scale height in hotter regions \citep[e.g.,][]{miller2008atmospheric}, such that they are ``puffed up'' and cover more solid angle on the sky. These hotter regions therefore contribute more to the observed net Doppler signal in HRCCS. The scale height effect is seen in \cite{kempton2012constraining} as a slight, increasing blueshift over transit and as slight ingress/egress differences. The effect there is not as dramatic as in planets with larger east--west limb asymmetry, such as WASP-76b \citep{west2016three, wardenier2021decomposing, savel2022no}.


With respect to equilibrium chemistry: because of the strong day--night contrasts in (ultra)hot Jupiter atmospheres, there exist strong spatial variations in temperature. The day--night contrasts result in east--west contrasts because the equatorial jet advects hot gas ahead of the substellar point to the evening limb and relatively cold gas from the antistellar point to the morning limb. Furthermore, as the planet rotates on its spin axis during transit, the hotter side of the planet progressively rotates into view, exacerbating these differences at egress. Ignoring all disequilibrium processes and scale height differences, there should therefore exist strong spatial variations in gas-phase atmospheric composition; at a given bulk composition, equilibrium chemistry implies variations in chemistry solely as a function of temperature and pressure. It is expected that asymmetries in transmission could hence vary as a function of temperature due to differences in chemistry alone. 

Chemical gradients are invoked to explain a number of observational datasets \citep[e.g.,][]{ehrenreich2020nightside, kesseli2021confirmation}. However, other temperature-dependent effects, such as the scale height effect, may instead be driving observed asymmetries. With this distinction in mind, it is prudent to consider the difference in strength between these two effects and whether one considerably outweighs the other.

\subsection{Asymmetry metric}

To quantify the asymmetry of chemical abundance in a planetary atmosphere, we construct a west--east asymmetry metric, $A_{\rm WE}$:

\begin{equation}\label{eq:awe}
\begin{split}
    A_{WE} = \frac{1}{C}\int_{\rm west} \log_{10}\bigg{(}{\int 
    n_{\alpha}(T, \,P)\, dl}\bigg{)} d\Omega \\- \frac{1}{C} \int_{\rm east} \log_{10}\bigg{(}\int n_{\alpha}(T, \,P)\, dl\bigg{)} d\Omega,
\end{split}
\end{equation}

\noindent where, for species $\alpha$, $n$ is the number density in an atmospheric cell, $d\Omega$ is the solid angle subtended by a given sky-projected radius--latitude cell, and there are $C$ total cells per limb. By equilibrium chemistry, $n$ is a function solely of temperature $T$ and pressure $P$ within a given cell in the modeled 3D atmosphere. For each 2D sky-projected radius--latitude cell, $dl$ is integrated along the line of sight through the planet's 3D modeled atmosphere. This metric takes into account regions of the planet outside the terminator (which impacts transmission spectra even at low resolution; e.g., \citealt{caldas2019effects}, \citealt{wardenier2022all}) by ray-striking through a 3D atmosphere.

$A_{\rm WE}$ essentially reduces to the difference in mean (log) abundance between the two limbs. The sign of this quantity encodes information about the asymmetry, as well: positive $A_{\rm WE}$ implies that the western limb is more abundant in a species, whereas negative $A_{\rm WE}$ implies that the eastern limb is more abundant in a species.

\subsection{Model atmospheres}
As of yet, we have remained agnostic to the model that generates the temperature--pressure structure and defines the grid cells for an $A_{\rm WE}$ calculation. Some of the most complex and physics-rich descriptions of 3D exoplanet temperature--pressure structures are given by general circulation models \citep[GCMs; e.g.,][]{showman2009atmospheric}. In this study, however, we seek to gain intuition for the basic scaling of asymmetry with planetary temperature (which drives the scale height and equilibrium chemistry gradients), and the added physical complexity of GCMs could add ``noise'' to this ``signal''---it would be difficult to isolate the effect of increasing planetary temperature alone. Furthermore, we here consider unphysical situations in order to determine the magnitude of the resulting difference with the correct physics. Finally, GCMs are very computationally expensive to run and have a number of free parameters to tune, and we here aim to explore a nontrivial grid of models over a representative range of parameter space.

We opt for a simple, parameterized approach instead of pursuing a full GCM description of our atmospheres for this specific experiment. Our model atmospheres have two parameters: a normalized east--west contrast $\Tilde{\Delta} T = (T_{\rm east} - T_{\rm west})/T_{\rm east}$ and an equilibrium temperature $T_{\rm eq}$. A normalized east--west contrast is a natural choice over an absolute east--west contrast for this work; namely, it prevents negative temperatures at low $T_{\rm eq}$, and it has physical meaning motivated by dynamical theory \citep[e.g.,][]{tan2019atmospheric}. In these models, the choice of $\Tilde{\Delta} T$ also uniquely enforces the east-west temperature differences. The limb-to-limb difference cannot exceed the day--night difference; based on the GCMs of \cite{tan2019atmospheric} and a set of phase curve observations \citep{parmentier2018exoplanet}, we do not expect a day--night contrast to exceed 0.6, so we hold our east--west contrast below this value.

Hence, we here sweep our parameterized atmospheric models in $\Tilde{\Delta} T$ from 0.1--0.6, in addition to sweeping in $T_{\rm eq}$ from 1000~K -- 4000~K. Each atmosphere is characterized by two isothermal temperature--pressure profiles. Defining

\begin{equation}
    \begin{split}
        T_{\rm east} = T_{\rm eq} + \Delta T/2 \\
        T_{\rm west} = T_{\rm eq} - \Delta T/2 \\
    \end{split}
\end{equation}

and noting that $\Delta T = T_{\rm east} - T_{\rm west}$, it therefore follows that

\begin{equation}
\begin{split}
    T_{\rm east} = \frac{T_{\rm eq}}{1 - \Tilde{\Delta} T/2} \\
    T_{\rm west} = \frac{T_{\rm eq}}{1 + \Tilde{\Delta} T/2}.
    \end{split}
\end{equation}

\begin{figure*}
    \gridline{\fig{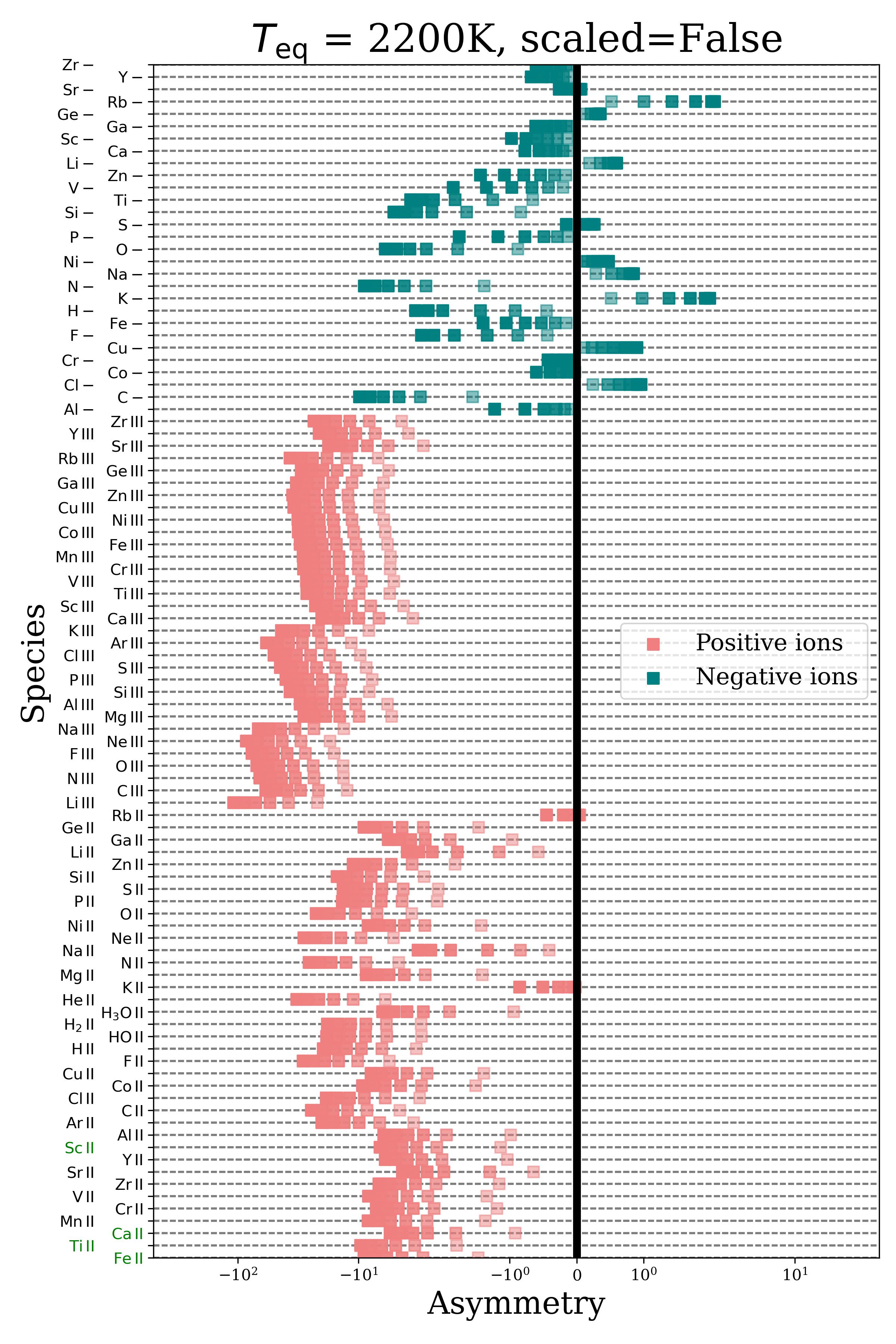}{0.5\textwidth}{(a)}
          \fig{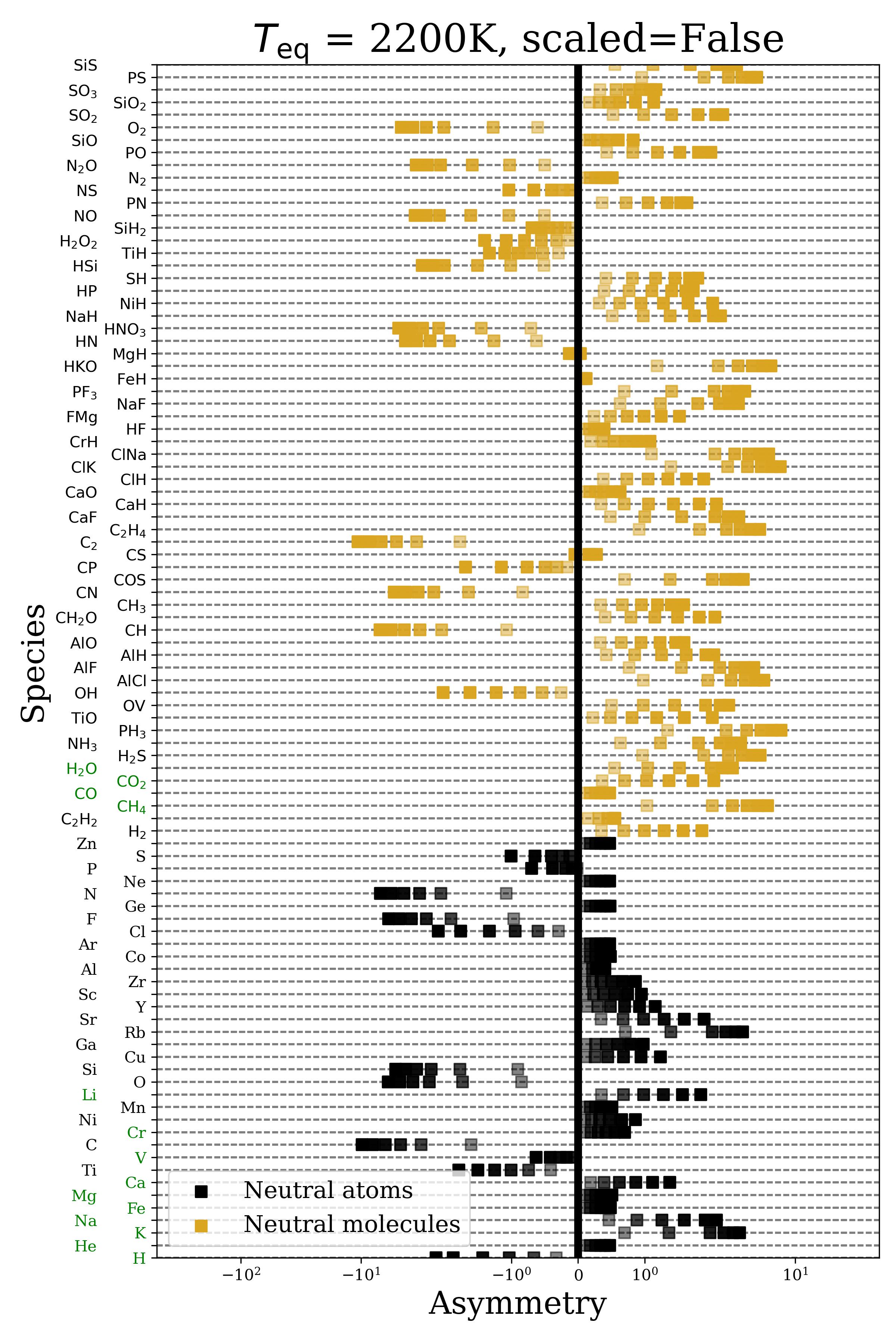}{0.5\textwidth}{(b)}
          }
\caption{Asymmetry (as defined in Equation~\ref{eq:awe}) of all chemical species considered in this study in our parameterized atmospheres at an equilibrium temperature of 2200~K. These models do not self-consistently inflate the hotter limb of the parameterized model (i.e., they do not observe the ``scale height effect''). The shading of each species represents the normalized temperature difference, $\Tilde{\Delta} T$, across the two limbs of our parameterized atmospheres; the lightest boxes have $\Tilde{\Delta} T=0.1$, whereas the darkest have $\Tilde{\Delta} T=0.6$. For illustrative purposes, we color in green tick marks for species with detections noted in \cite{guillot2022giant} (and including the recent $\rm CO_2$ detection; \citealt{early2022identification}). We also draw a vertical line denoting 0 asymmetry. Without taking the scale height effect into account, positive ions form much more predominantly on the warmer limb (i.e., have negative asymmetry) than other species and reach the greatest asymmetry values.
\label{fig:asymm_all_3000}}
\end{figure*}

\begin{figure*}
    \gridline{\fig{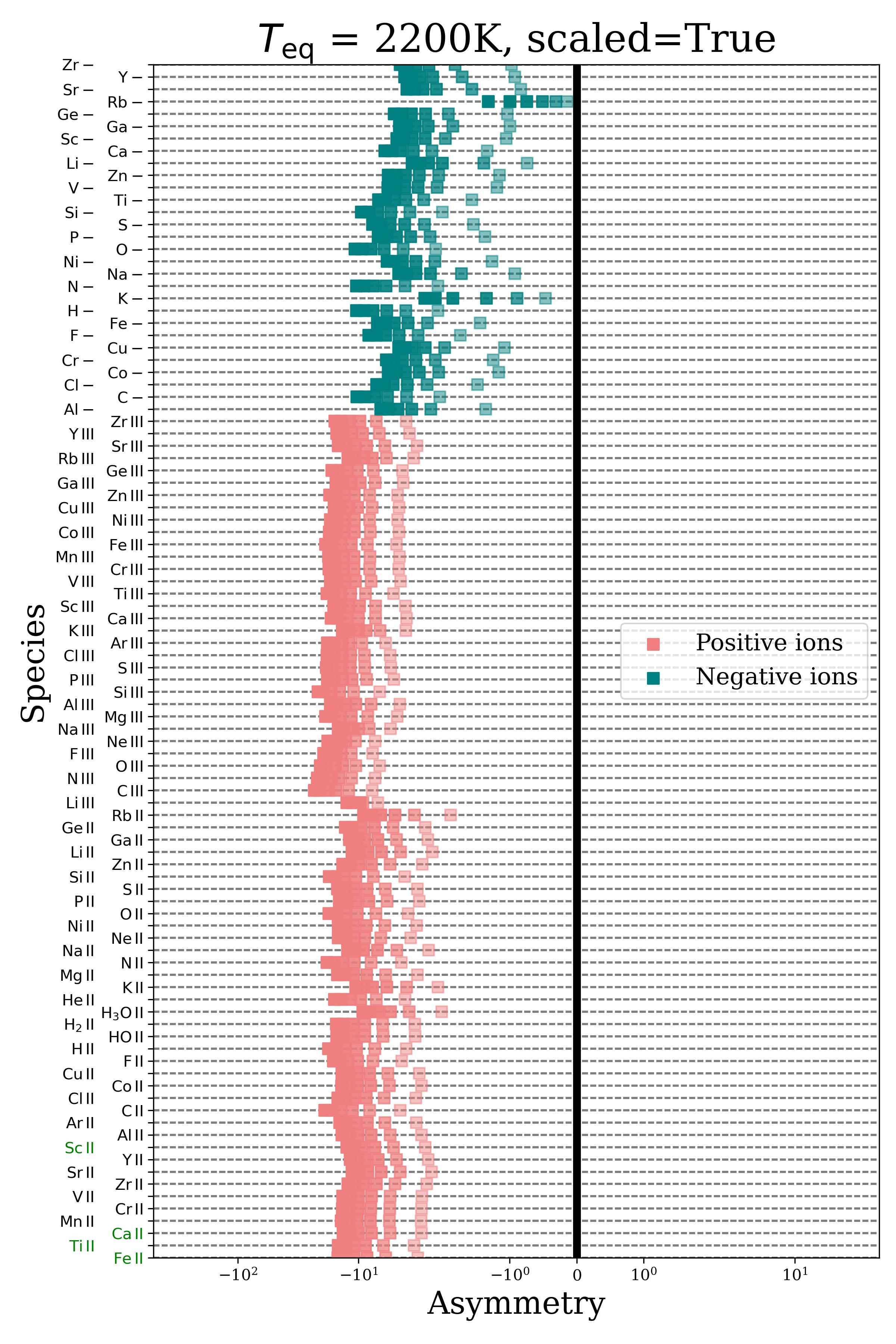}{0.5\textwidth}{(a)}
          \fig{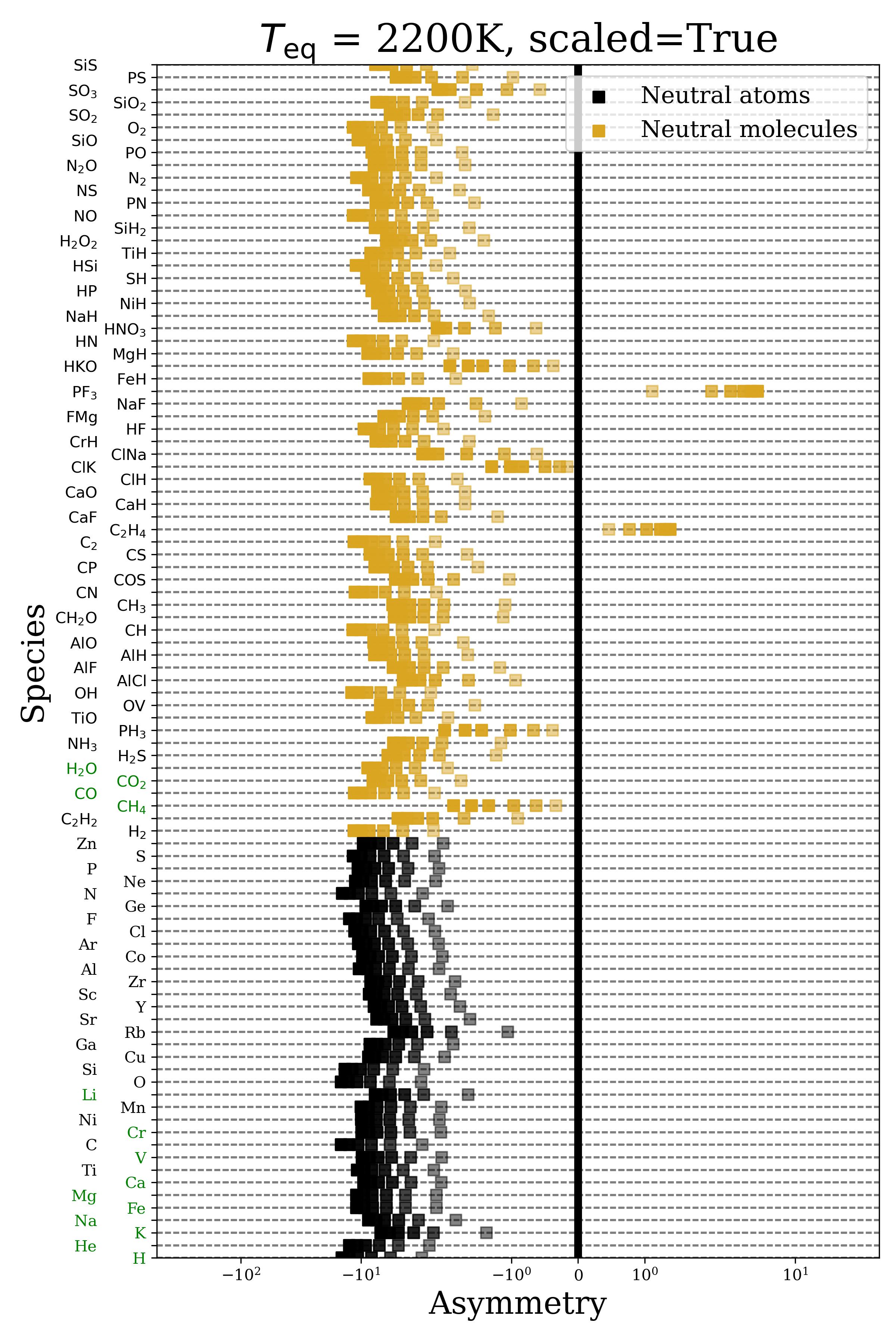}{0.5\textwidth}{(b)}
          }
\caption{Similar to Figure~\ref{fig:asymm_all_3000}, but now including the scale height effect (inflating the hotter limb in our parameterized models). Now, all species have asymmetries that favor the hotter limb (negative asymmetry)---simply because the hotter limb subtends more solid angle on the sky. However, there still exists inter-species variability in asymmetry, implying that the scale height effect does not entirely swamp genuine differences in equilibrium chemistry across limbs. Furthermore, negative ions still have larger asymmetries than positive ions or neutral species.
\label{fig:asymm_all_3000_scaled}}
\end{figure*}

\begin{figure*}
\includegraphics[scale=0.5]{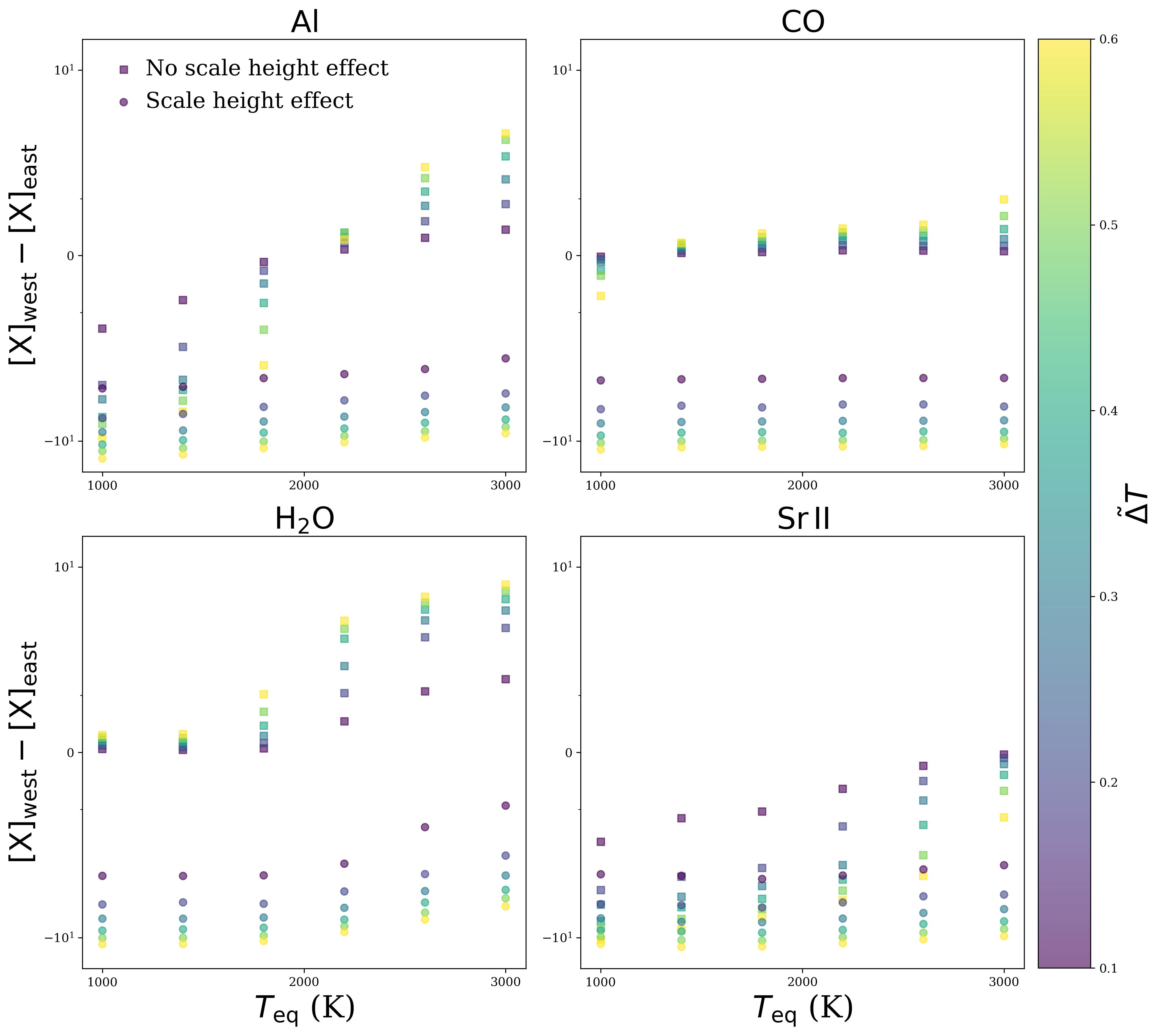}
\centering
\caption{Asymmetry (per Equation~\ref{eq:awe}) for Sr~II, Fe, $\rm H_2O$, and CO in our parameterized atmospheres. Our grid sweeps over equilibrium equilibrium temperature and normalized temperature difference across limbs, and includes models that observe the scale height effect (circles) and do not (squares). We find that species with strong temperature-dependent abundances (e.g., Sr~II) are less dominated by the scale height effect than species with weaker temperature-dependent abundances.
\label{fig:asymm_fe_sr}}
\end{figure*}

With the substellar longitude at $0\degr$, all cells with a longitude $\phi < 180\degr$---the warmer evening limb---are assigned temperature $T_{\rm east}$. Conversely, all cells with a longitude $\phi > 180\degr$---the cooler morning limb---are assigned temperature $T_{\rm west}$. Pressures in the atmosphere run as low as 1 $\mu$bar, as one of the benefits of HRCCS is that it can probe low pressures such as these \citep[e.g.,][]{kempton2014high, gandhi2020seeing, hood2020prospects}. The bottom of the atmosphere is set at 0.5 bar; our previous 3D forward models run in \cite{savel2022no} across the optical and near-infrared indicate that for our test case of WASP-76b \citep{west2016three}, this region is the deepest that can be probed given the expected continuum opacity. The parameterized modeled atmospheres in this study have no set wind fields, as in our models (motivated by and assuming chemical equilibrium), winds do not control $A_{\rm WE}$---only the chemical abundance of a given cell does.

We calculate $A_{\rm WE}$ to assess the relative strength of the scale height and equilibrium chemistry effects. To infer the strength of the scale height effect, we construct pairs of model atmospheres. In each pair, one atmosphere is constructed self-consistently: pressure falls off per hydrostatic equilibrium, with the scale height set by the temperature on either limb. For the models here, we hold composition constant across both limbs, thereby holding $\mu$ constant at 2.36 \citep[appropriate for a solar-composition gas dominated by molecular $\rm H_2$;][]{kempton2012constraining}. See Section~\ref{sec:asymmetry_application} for a discussion of this caveat. The other atmosphere in the pair is constructed on the same pressure grid as the western limb at all longitudes. That is, the eastern limb is not simulated as inflated compared to the western limb---removing the scale height effect from the projected model atmosphere in transmission.

\subsection{Equilibrium chemistry}\label{sec:asymmetry_chemistry}

To calculate the number densities of our species in each modeled atmospheric cell ($n_{\alpha}$), we construct a grid in temperature--pressure space using the \texttt{FastChem} equilibrium chemistry code \citep{stock2018fastchem} and interpolate the grid based on local atmospheric cell temperature and pressure. We initialize the code with solar abundances from \cite{lodders2003solar}. Our chemistry code does not explicitly include any condensation or cloud-formation processes.

Even disregarding questions of species detectability in HRCCS data, it is worth considering that not all species with \texttt{FastChem} thermochemical data have freely available opacity data. With this constraint in mind, we restrict our $A_{\rm WE}$ molecule calculations to molecules with opacity data available on ExoMol,\footnote{\url{https://www.exomol.com/}} a popular opacity database for exoplanet atmosphere modeling.

\subsection{Asymmetry metric: application}\label{sec:asymmetry_application}

We calculate $A_{\rm WE}$ for our grid of parameterized atmospheres. Disregarding the scale height effect, we find that positive ions tend to form preferentially on the hotter limb of our models at an equilibrium temperature of 2200~K (Figure~\ref{fig:asymm_all_3000}). This is expected, as thermal ionization should increase the abundance of positive ions at higher temperatures. Furthermore, larger east--west temperature asymmetries lead to larger abundance asymmetries.

Including the scale height effect increases the asymmetry for neutral atoms and molecules, as can be seen by comparing the right-hand sides of Figures~\ref{fig:asymm_all_3000}--\ref{fig:asymm_all_3000_scaled}. Furthermore, there is more homogeneity across the $A_{\rm WE}$ values across positive ions, negative ions, neutral atoms, and neutral molecules (Figure~\ref{fig:asymm_all_3000_scaled}). In particular, while higher $\Tilde{\Delta} T$ still implies higher absolute asymmetry in neutral species, the scale height effect makes it such that the warmer limb almost always has higher projected asymmetry.

It is therefore clear that the scale height effect strongly tamps down genuine variation in species abundance due to equilibrium chemistry. However, the fact that inter-species variation in asymmetry remains implies that this variation in abundance is not completely washed out by the scale height effect; if the scale height effect truly and fully dominated, all species would have the same $A_{\rm WE}$ value.


When considering individual species more closely, we find that certain species are particularly differentially affected by the scale height effect. For example, Figure~\ref{fig:asymm_fe_sr} shows that there is a stark difference in whether the scale height effect is included for Fe. However, this is not as much the case for, e.g., Sr~II. The meaning behind this result is evident in the equilibrium abundance calculations of Fe and Sr~II: Fe is less sensitive to temperature variations than Sr~II. This result is expected, as the onset of Sr~II is determined by the temperature at which Sr~I can be effectively ionized. This is generally the case for positive ions---the temperature effect on chemical abundance wins out over the scale height effect, as seen by the left-hand sides of Figures~\ref{fig:asymm_all_3000}--\ref{fig:asymm_all_3000_scaled}. Physically, this behavior is because the Saha equation is more strongly dependent on temperature than most chemical equilibrium reaction rates.

The results of this experiment indicate that the most temperature-sensitive species are strongly influenced by both abundance changes and scale height differences. Conversely, to isolate the scale height effect, it would be therefore useful to consider a species with very weakly temperature-dependent abundance; in this case, if a strong asymmetry were detected, it could be attributed to a scale height effect (or other non-equilibrium chemistry or physics). We explore this idea further in Section~\ref{sec:diagnostics}.

Note that this approach, aside from its simplified temperature--pressure structure, does not account for a variety of physics. Namely, it does not include the effects of hydrogen dissociation and recombination that occurs in the ultra-hot Jupiter regime \citep{tan2019atmospheric}. Inclusion of this physics would serve to decrease the mean molecular weight in the atmosphere, increasing the scale height for the hotter, eastern limb, thereby amplifying the observed asymmetry. Additionally, at the lower-temperature end, we did not include the effects of certain species being sequestered into clouds (e.g., silicate clouds). We will model the Doppler shift impact of optically thick clouds in Section~\ref{sec:clouds less blueshifting}. \addressresponse{Finally, our approach does not include disequilibrium effects (e.g., vertical / horizontal mixing) that may alter asymmetries. Therefore, the results shown here motivate asymmetries due to equilibrium chemistry alone, which we expect to be a first-order driver of asymmetry; disequilibrium chemistry is not expected to be significant in the ultrahot Jupiter regime \citep[e.g.,][]{tsai2021comparative}.} 

We further did not include the effect of temperature- and pressure-dependent opacities. At the spectrum level, a temperature asymmetry would be exaggerated by the fact that, e.g,. Fe absorbs more on the hotter limb than the colder limb because its opacity increases with temperature. This would mean that the detected net Doppler shift is even more strongly weighted to the hotter limb.

Despite these limitations in our modeling, the trends listed above should hold to first order and provide intuition about the relative strengths of two potential drivers of asymmetry in exoplanet atmospheres. Broadly, it holds that the scale height effect appears to dominate in general, but relative differences in abundances of species as a function of temperature still matter. Given the limitations of simple models, we will move on to more self-consistent atmospheric modeling in the following sections.

\section{Selected diagnostics}\label{sec:diagnostics}

\subsection{Diagnostics for specific mechanisms}
Per Section~\ref{sec:drivers}, even differentiating between two drivers of asymmetry in exoplanet atmospheres is nontrivial. Drivers can compete to varying degrees to produce a similar result: an asymmetric trend in net Doppler shifts in HRCCS.

However, by exploiting nuances in the HRCCS Doppler shift signal and by independent means, it may be possible to disentangle even drivers that produce similar effects. Table~\ref{table:tests} lists example drivers of asymmetries in HRCCS and how they might be diagnosed. The associated works listed in the table may not directly propose these diagnostics, but at minimum they provide foundational material for them. 

Of course, exhibiting a single diagnostic does not not mean that a given physical mechanism is in play. Other mechanisms could surely be present, and uniquely constraining a single mechanism as dominant would require ruling out the others, as well. For instance, both day--night winds and morning limb condensation could result in a net blueshifted CCF. But if, for example, a nightside temperature were derived from a phase curve that was far too hot for any known condensate to form, then day--night winds would be much preferred to condensation as a physical solution. Together, collections of diagnostics are hence able to test the dominance of individual mechanisms.

In the following sections, we explore a few tests for specific physical mechanisms of asymmetry: using CO as a baseline molecule to identify the scale height effect and tracking the blueshifts of multiple species to identify the presence of clouds. We furthermore evaluate the effectiveness of diagnostics that may be used to evaluate a number of different mechanisms: averaging HRCCS data into two phase bins and using finely phase-resolved HRCCS data. We additionally show how these diagnostics can further motivate or rule out ``toy models'' that at first may appear convincing.

\subsubsection{CO as a baseline molecule}\label{sec:co_baseline}
We have demonstrated (Section~\ref{sec:asymmetry_application}) that species with strongly temperature-dependent abundances are the least susceptible to the scale height effect. Conversely, observing a species with very \textit{weak} temperature-dependent abundance could indicate whether the scale height effect is in play. 

Consider CO. In Figure~\ref{fig:asymm_fe_sr}, its $A_{\rm WE}$ values are clustered around 0 without the scale height effect, with relatively weak dependence on $\Tilde{\Delta} T$. However, CO's $A_{\rm WE}$ values are strongly negative when the scale height effect is included. We propose using CO as a tracer of the scale height (and other chemistry-unrelated) effects.

\begin{figure}
    \centering
    \includegraphics[scale=0.38]{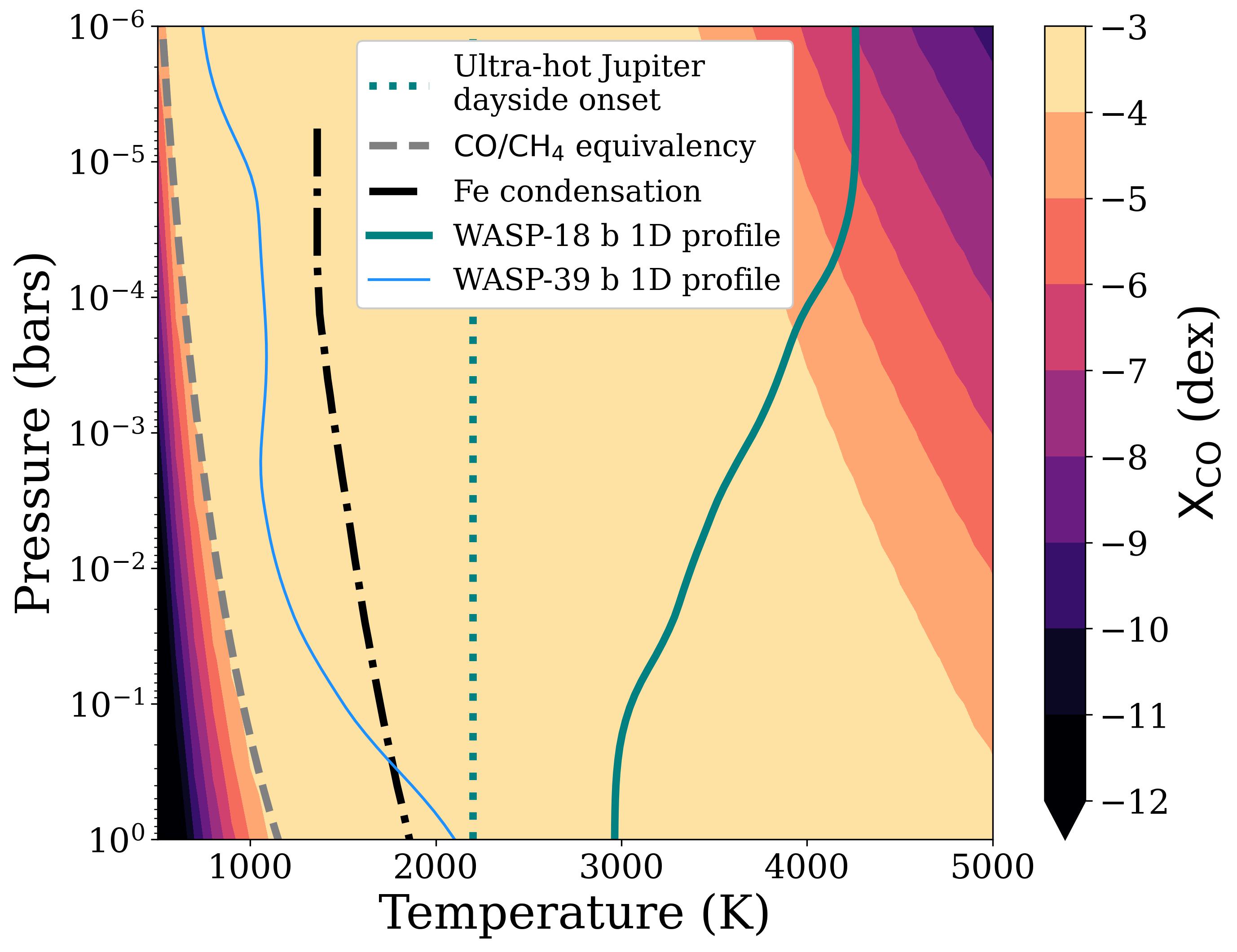}
    \caption{Volume mixing ratio of CO as a function of pressure and temperature as calculated by \texttt{FastChem}. Overplotted are the onset of ultra-hot Jupiters \citep[as defined by their dayside temperature;][]{parmentier2018thermal}, the CO/$\rm CH_4$ equivalency curve from \cite{visscher2012chemical} as a function of pressure, the Fe condensation curve from \cite{mbarek2016clouds}, and 1D temperature--pressure profiles for a hot Jupiter (WASP-39b) and an ultra-hot Jupiter (WASP-18b) as computed with \texttt{HELIOS} \citep{malik2017helios}. Both the condensation curve and the equivalency curve are computed at solar metallicity. Considering the regime of ultra-hot Jupiter atmospheres, CO is a relatively stable chemical species.}
    \label{fig:co_baseline}
\end{figure}

As shown in Figure~\ref{fig:co_baseline}, the abundance of CO is relatively stable between 1000~K and 3500~K. \cite{beltz2022emission} note that this stability holds over the temperature--pressure range of the observable atmosphere of the ultra-hot Jupiter WASP-76 b. Indeed, this feature remains true over the general temperature--pressure range of ultra-hot Jupiters. For illustrative purposes, we calculate the 1D temperature--pressure profiles of a hot Jupiter (WASP-39b; \citealt{faedi2011wasp}) and an ultra-hot Jupiter (WASP-18b; \citealt{hellier2009orbital}). These profiles, calculated with the \texttt{HELIOS} 1D radiative-convective model (with full heat redistribution), indicate that the observable atmosphere for these planets is largely within a region of near-constant CO mixing ratio. The stability of CO is due to three factors: its strong chemical bonding, its lack of participation in gas-phase chemical reactions, and its lack of condensation.

Since the strong triple bond of CO makes it difficult to thermally dissociate, CO remains stable at temperatures that would dissociate molecules with weaker bonds, such as $\rm H_2O$ \citep{parmentier2018thermal}, which has two single bonds. Beyond roughly 3500~K, even the triple bond becomes susceptible to thermal dissociation; hence, the few exoplanets with significant portions of their atmosphere hotter than this temperature \citep[e.g., KELT-9b, with $T_{\rm eq} \approx 4050$~K;][]{gaudi2017giant} would likely exhibit spatial variation in CO abundance. Most ultra-hot Jupiters, though, should fall shy of this regime. Furthermore, the high photoionization threshold of CO \citep[relative to, e.g., $\rm H_2O$;][]{heays2017photodissociation} means that it is not commonly photodissociated \citep{van1988photodissociation}. Even when it is photodissociated, recyclying pathways exist in hot Jupiters that can replenish CO abundance, keeping it near equilibrium abundance even inclusive of photochemistry \citep{moses2011disequilibrium}. Hence, the assumption of non-dissociation of CO is reasonably justified across much of the ultra-hot Jupiter population. 

Additionally, CO does not commonly participate in thermochemical reactions and is the dominant carbon carrier in our temperature--pressure range of interest. While at lower temperatures the dominant carbon carrier becomes $\rm CH_4$, the ultra-hot Jupiter regime is squarely beyond the CO/$\rm CH_4$ equivalency curve \citep[Figure~\ref{fig:co_baseline};][]{visscher2012chemical}. Therefore, even aside from thermal dissociation, CO should not participate in gas-phase thermochemistry that would alter its abundance.

Finally, CO does not form any high-temperature condensates expected in ultra-hot Jupiter atmospheres. The condensation temperature of CO \citep[$\approx$80~K at 1~bar;][]{lide2006crc,fray2009sublimation} is far below the temperature--pressure range of ultra-hot Jupiters. This quality makes CO a less complicated tracer of, e.g., atmospheric dynamics than species that do condense in this region of parameter space, such as Fe, Mg, or Mn \citep{mbarek2016clouds}. Therefore, while the calculations of Figure~\ref{fig:co_baseline} do not include gas-phase condensation, the resultant spatial constancy of CO should still be robust even when condensation is considered. CO is thus a more straightforward molecule to model than other, condensing species, as it does not participate in the complex microphysics of condensation and cloud formation \citep[see, e.g.,][]{gao2021aerosols}.

Beyond its spatial uniformity, there are further observational reasons that CO is an appealing species to target. 
Namely, CO has very strong spectroscopic bands placed across the infrared wavelength range \citep[e.g.,][]{li2015rovibrational} that do not overlap with other strong absorbers and are relatively well understood \citep{li2015rovibrational}. Additionally, the high cosmic abundance of C and O \citep{lodders2003solar} means that, unlike many of the species in the previous section, CO is readily detectable \citep[and has been become a standard detection in HRCCS;][]{snellen2010orbital, de2013detection,rodler2013detection, brogi2014carbon, brogi2016rotation, flowers2019high, giacobbe2021five, line2021solar,pelletier2021water, zhang202112co,Guilluy2022five, van2022carbon}.


Given its stability and observational advantage, we propose that CO can be used as a faithful tracer of the atmosphere---whether it is inflated in some regions, what its wind profile is, whether regions are blocked by clouds, etc. In turn, CO may then be leveraged to better motivate sources of asymmetry that affect other species. While other species with low $A_{\rm WE}$ in Figure~\ref{fig:asymm_all_3000} (e.g., He, Fe, MgH, Rb II) would also appear to be good candidates for baseline species, these species are either largely spectroscopically inactive, have variable abundance over broader temperature--pressure ranges, or can condense. A caveat to the above is that while CO is a faithful longitudinal tracer, it is not an unbiased \textit{radial} tracer (as seen in Figure~\ref{fig:co_baseline}). As with all chemical species, CO has its own balance between deep and strong lines that depends on the waveband considered (see, e.g., Section~\ref{sec:phase_resolved_nir}). Therefore, the net CO Doppler signal does not uniformly weight the wind profile across all altitudes. Again, this is a bias inherent to all chemical species.





\subsubsection{A decreasing blueshift test for clouds}\label{sec:clouds less blueshifting}
As noted in Table~\ref{table:tests}, clouds may introduce strong asymmetry into HRCCS data. \cite{savel2022no} demonstrated that gray, optically thick clouds produce stronger blueshifts in the Doppler shift signal of WASP-76b than the blueshifts in clear models, also changing the trend of Doppler shift with phase. But, again as shown in Table~\ref{table:tests}, these changes at the Doppler shift level are not sufficient to uniquely identify clouds as the driver of an observed asymmetry. Combinations of observable quantities that would uniquely identify clouds as the source of observed HRCCS asymmetry are therefore necessary.

To devise such a test, we investigate in this work a limiting-case cloudy model. As in \cite{savel2022no}, we construct gray, optically thick, post-processed clouds in our 3D ray-striking code. We here make another assumption, though: that the clouds are confined to the cooler, \addressresponse{morning limb}, as opposed to having a distribution dictated by a specific species' condensation curve. This distribution is based on planetary longitude (between longitudes of $180\degr$ and $360\degr$). This approach is motivated by the results of \cite{roman2021clouds}, who found that a subset of cloudy GCMs exhibited a cloud distribution strongly favoring the western limb.\footnote{These GCMs produced clouds on a temperature--pressure basis, and did not model clouds as tracers. Therefore, they do not capture potential disequilibrium cloud transport (e.g., as done in \citealt{komacek2022patchy}), which may alter the degree of patchiness within the cloud deck.} Our approach benefits from providing limiting-case intuition for how cloudiness affects Doppler shift signals while avoiding the complex questions of how clouds form and which species contribute the most opacity \citep{gao2021aerosols, gao2021universal}.

Briefly, our modeling methodology is as follows:

\begin{enumerate}
    \item Double-gray, two-stream GCM. GCMs such as this one solve the primitive equations of meteorology, which are a reduced form of the Navier-Stokes equations solved on a spherical, rotating sphere with a set of simplifying assumptions.\footnote{These assumptions are 1) local hydrostatic equilibrium, such that vertical motions are caused purely by the convergence and divergence of horizontal flow, 2) the ``traditional approximation,'' which removes the vertical coordinate from the Coriolis effect, and 3) a thin atmosphere.} The output of these models is temperature, pressure, and wind velocity as a function of latitude, longitude, and altitude. We use the GCM that was shown to best fit the \cite{ehrenreich2020nightside} WASP-76b data in \cite{savel2022no}. 
    \item Equilibrium chemistry with \texttt{FastChem}. As in Section~\ref{sec:asymmetry_chemistry}, we interpolate a model grid of chemistry to determine local abundances of a number of chemical species as determined by temperature and pressure conditions of the GCM output.
    \item Ray-striking radiative transfer. Using a code modified from \cite{kempton2012constraining} \citep[as detailed in][]{savel2022no}, we compute the high-resolution absorption spectrum of our planetary atmosphere by calculating the net absorption of stellar light along lines of sight through our GCM output. This absorption is calculated inclusive of net motions along the lines of sight from atmospheric winds and rotation, inducing Doppler shifts relative to that of a static atmosphere's spectrum. Limb-darkening is calculated with a quadratic limb-darkening law in the observable planetary atmosphere and with the \texttt{batman} code \citep{kreidberg2015batman} for the portion of the star blocked by the optically thick planetary interior.
\end{enumerate}

Given its increasing utility as a benchmark planet for HRCCS studies \addressresponse{\citep[e.g.,][]{ehrenreich2020nightside, kesseli2021confirmation, landman2021OH, seidel2021into, wardenier2021decomposing, kesseli2022atomic, sanchez2022searching}}, we model the ultra-hot Jupiter WASP-76b \citep{west2016three}. We calculate 25 spectra inclusive of Doppler effects equally spaced in phase from the beginning to end of transit. For our cross-correlation template, $\mathcal{T}$, we use a model that does not include Doppler effects.

\begin{figure}
    \centering
    \includegraphics[scale=0.25]{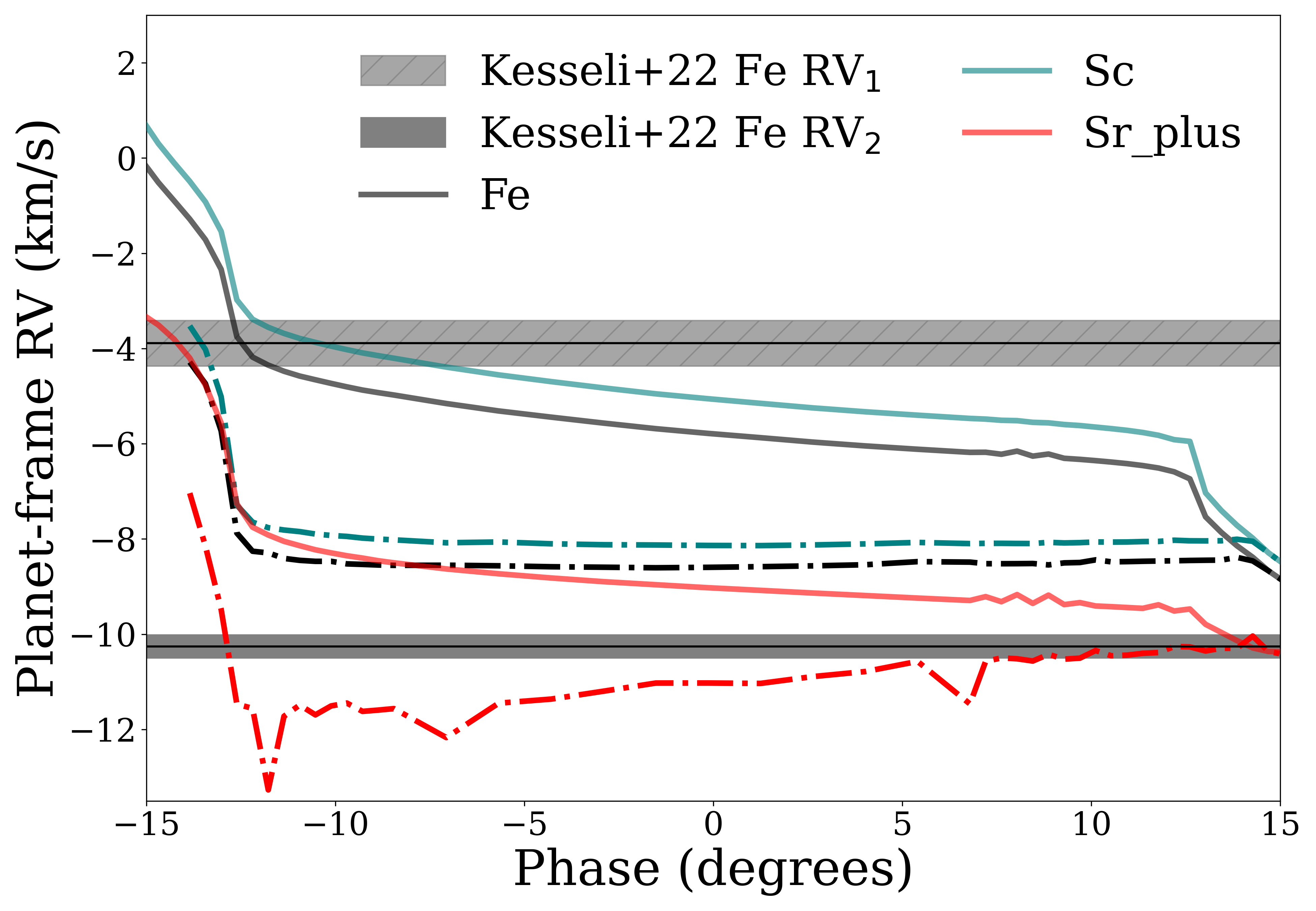}
    \caption{Atmospheric Doppler shifts, which should remain in the HRCCS signal after the orbital motion is subtracted, as a function of orbital phase for our forward models. Shown are representative species that span Doppler shifts and are noted as potentially observable by \cite{kesseli2022atomic}: Fe, Sr~II, and Sc. Cloud-free models are represented with solid lines, whereas models with fully cloudy morning limbs are represented with dashed lines. \addressresponse{The first half of transit (RV$_1$) and second half of transit (RV$_2$) Doppler shifts for Fe from \cite{kesseli2022atomic} are overplotted as horizontal lines, with width determined by observational errors. Our} cloudy models are much more strongly blueshifted than their cloud-free counterparts, become less blueshifted over transit, and do not have significant CCF peaks at early phases.}
    \label{fig:cloudy_doppler}
\end{figure}

We then cross-correlate our template against our calculated spectrum, $y$:

\begin{equation}
    c(v) = \sum_{i=0}^{N} y_i(\lambda)\mathcal{T}_i(v, \lambda),
\end{equation}

\noindent where the mask or template is Doppler-shifted by velocity $v$ and interpolated onto the wavelength grid, $\lambda$, of $y$ for summing. Our CCF is computed on a grid of velocities from $-250$~km\,s$^{-1}$ and 250~km\,s$^{-1}$ with a step of 1~km\,s$^{-1}$. The final net planet-frame Doppler shift is calculated by fitting a Gaussian to the peak of the CCF.

The results of our experiment are shown in Figure~\ref{fig:cloudy_doppler}. When we allow clouds to extend over the entire morning limb, note that all species become less blueshifted over time. Because the limb that is rotating away from the observer (the ``receding limb'') is entirely blocked off by clouds, there is no wavelength-dependent absorption for that limb. Therefore, the contribution of redshifting from solid-body rotation on the receding limb is not present---the only Doppler shift contributions are from evening limb rotation and evening limb winds, which are generally in the same direction as the rotation. Hence, there are much stronger blueshifts at earlier phases than in the clear models.

However, at later phases, the non-cloudy regions of the atmosphere rotate into the receding limb, thereby contributing some rotational redshift to the net Doppler shift signal.\footnote{The degree of rotation during transit varies as a function of semimajor axis and host star radius, and hence from planet to planet. While we only model WASP-76b, other planets also have large (e.g., compared to the angles probed by transmission spectroscopy) rotations during transit \citep{wardenier2022all}.} At the earliest phases, the cloudy models do not have enough wavelength-dependent absorption to produce a significant CCF peak. Notably, all species follow this trend, as the blocking of clouds as modeled here is wavelength-independent and altitude-independent. This behavior is shown in Figure~\ref{fig:cloudy_doppler} for Fe, Sc, and Sr~II---all species identified in \cite{kesseli2022atomic} has having high potential observability for ultra-hot Jupiters.

\addressresponse{From Figure~\ref{fig:cloudy_doppler}, it is also apparent that the cloud-driven trend of decreasing blueshift in phase is not matched by the observations of \cite{kesseli2022atomic}. As found in \cite{savel2022no} in comparison to the \cite{ehrenreich2020nightside} data, while the absolute magnitude of the cloudy model's Doppler shift better match the data than the clear model's, the cloudy model trend over Doppler shift is not matched by the data. In sum, this limiting-case model of opaque, morning limb clouds does not appear to be a first-order effect driving existing observational trends. This does not necessarily mean that clouds are not the driving factor behind limb asymmetries; it may simply be that a more physically motivated model for partial cloud coverage of the limb could fit the available data better.}

Also of note in Figure~\ref{fig:cloudy_doppler} is that the egress signatures of the clear and cloudy models are quite distinct. Near a phase of roughly 14 degrees, the clear model produces a sharp change in Doppler shift for all species as the leading (rotationally redshifted) limb begins to leave the stellar disk. This sharply blueshifting behavior continues to the end of egress, until the last sliver of the trailing (rotationally blueshifted) limb has left the stellar disk as well. In the cloudy case, however, the leading limb leaving the stellar disk has no effect, as it is fully cloudy. While this effect is evident in these models, it may be less evident in observations, which naturally cannot finely sample ingress and egress phases.

\subsection{Phase bins}\label{sec:phase bins}
We have thus far examined drivers of asymmetry and potential diagnostics of specific mechanisms. Next, we will evaluate a few HRCCS data types to determine how robust they are and their potential ability to constrain a number of different physical mechanisms that give rise to HRCCS asymmetry.

The first of these data types is HRCCS Doppler shifts that are binned in phase. A substantial fraction of HRCCS studies present detections and Doppler shifts integrated over the entirety of transit \citep[e.g.,][]{giacobbe2021five}. This approach maximizes detection SNR, which may be necessary for a given set of observations (e.g., because of a low-resolution spectrograph, small telescope aperture, faint star, low species abundance, low number of absorption lines, or weak intrinsic absorption line strengths). While it is possible to reveal aspects of limb asymmetry with this approach, especially when comparing detections of multiple species to one another, phase-resolving the transit (and observing isolated ingresses and egresses when possible) will certainly give a more direct probe of east--west asymmetries. Binning HRCCS data in phase across transit may provide a desirable balance between revealing asymmetry and maintaining high SNR.

We seek to address this question by phase-binning modeled Doppler shifts to examine its biases with respect to the underlying model. We follow this experiment with a comparison to the phase-binned observations of \cite{kesseli2022atomic}.

\subsubsection{Theoretical phase binning}\label{sec:theoretical phase binning}
We average our phase-resolved calculations into two bins: the first and second half of transit. Once our CCFs are calculated we average them in phase to effectively reduce our data to two single bins: the first half of transit and the second half of transit. We make versions of these two half-transit bins that include or exclude the ingress and egress phases (when the planet is only partially occulting the star).

\begin{figure}
    \centering
    \includegraphics[scale=0.58]{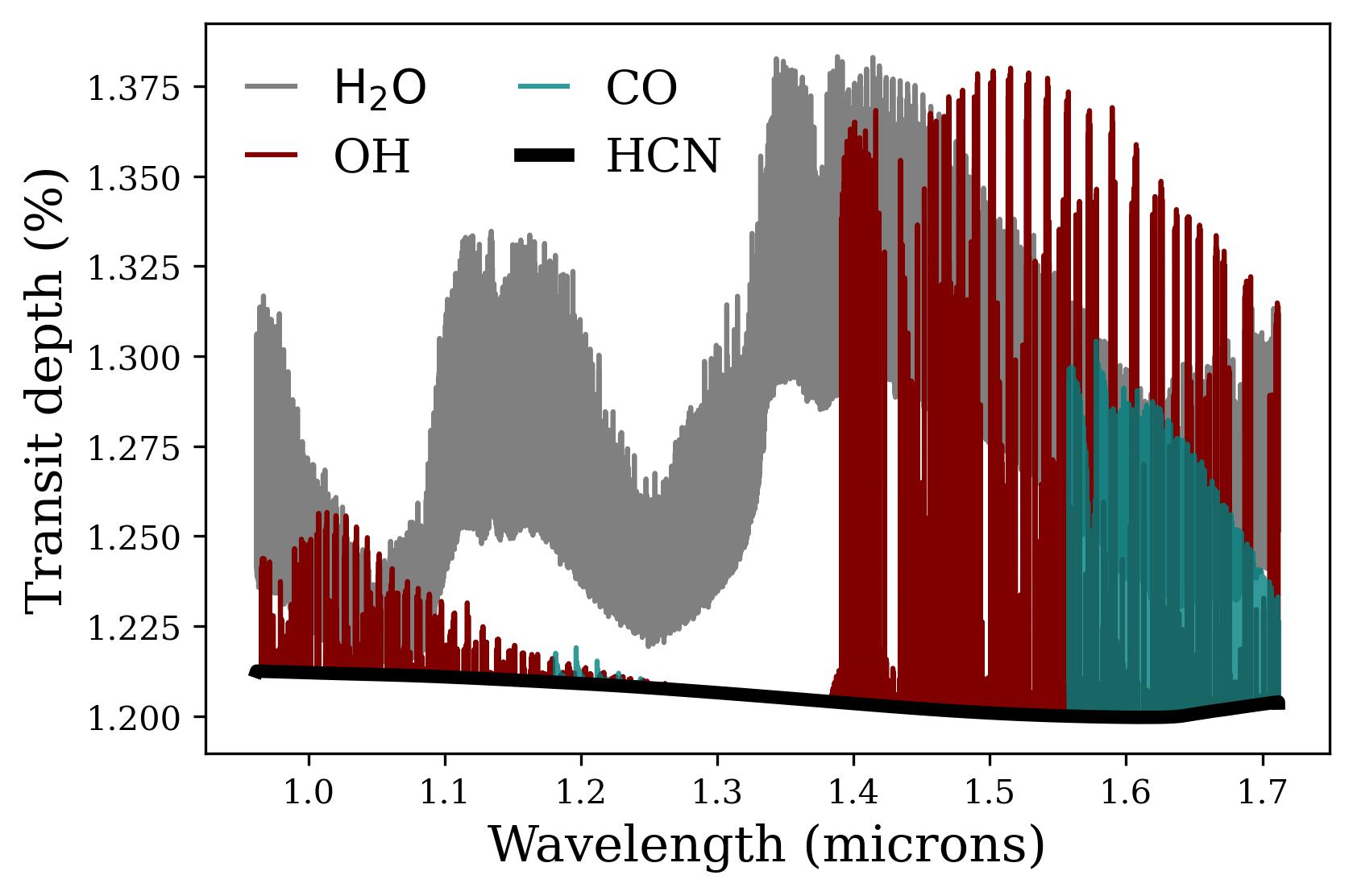}
    \caption{Single-species (OH, CO, $\rm H_2O$, HCN) 3D forward-modeled spectra of WASP-76b. These spectra are simulated over the CARMENES waveband and resolution. Doppler effects are not included in these spectra, which are modeled at center of transit. $\rm H_2O$ is the dominant absorber in this bandpass, followed by OH. HCN exhibits no spectral features above the continuum for WASP-76b in this bandpass.}
\label{fig:nir_spectrum}
\end{figure}

Motivated by recent detections in the near-infrared \citep{landman2021OH, sanchez2022searching}, we search for absorption from various molecules\footnote{We use the MoLLIST \citep{brooke2016line}, POKAZATEL \citep{polyansky2018exomol}, and Li2015 \citep{li2015rovibrational} linelists for OH, $\rm H_2O$, and CO, respectively.} in our models, focusing on the CARMENES \citep{quirrenbach2014carmenes} wavelength range and resolution for direct comparison against observational results using that instrument. Of these molecules, we find that OH, $\rm H_2O$, and CO produce significant absorption over the modeled wavelength range, with OH and $\rm H_2O$ producing the strongest features (Figure~\ref{fig:nir_spectrum}). We find that HCN does not produce any noticeable absorption under the assumption of chemical equilibrium and solar composition, implying either more exotic chemistry for WASP-76b's atmosphere (i.e., photochemistry or non-solar abundances; \citealt{moses2012chemical}), or that the detection of HCN in this atmosphere \citep{sanchez2022searching} was spurious \citep[perhaps due to the nature of the HCN opacity function;][]{zhang2020platon}. \addressresponse{We furthermore find a moderate ($\approx 4$~km\,s$^{-1}$) increase in blueshift for our modeled $\rm H_2O$. While this increase in blueshift is commensurate with the increase in blueshift described for $\rm H_2O$ in \cite{sanchez2022searching}, we are once again unable to match the high reported velocities (here -14.3~km\,s$^{-1}$) with our self-consistent forward models.}

\begin{figure*}
    \centering
    \includegraphics[scale=0.6, trim=200 60 200 40]{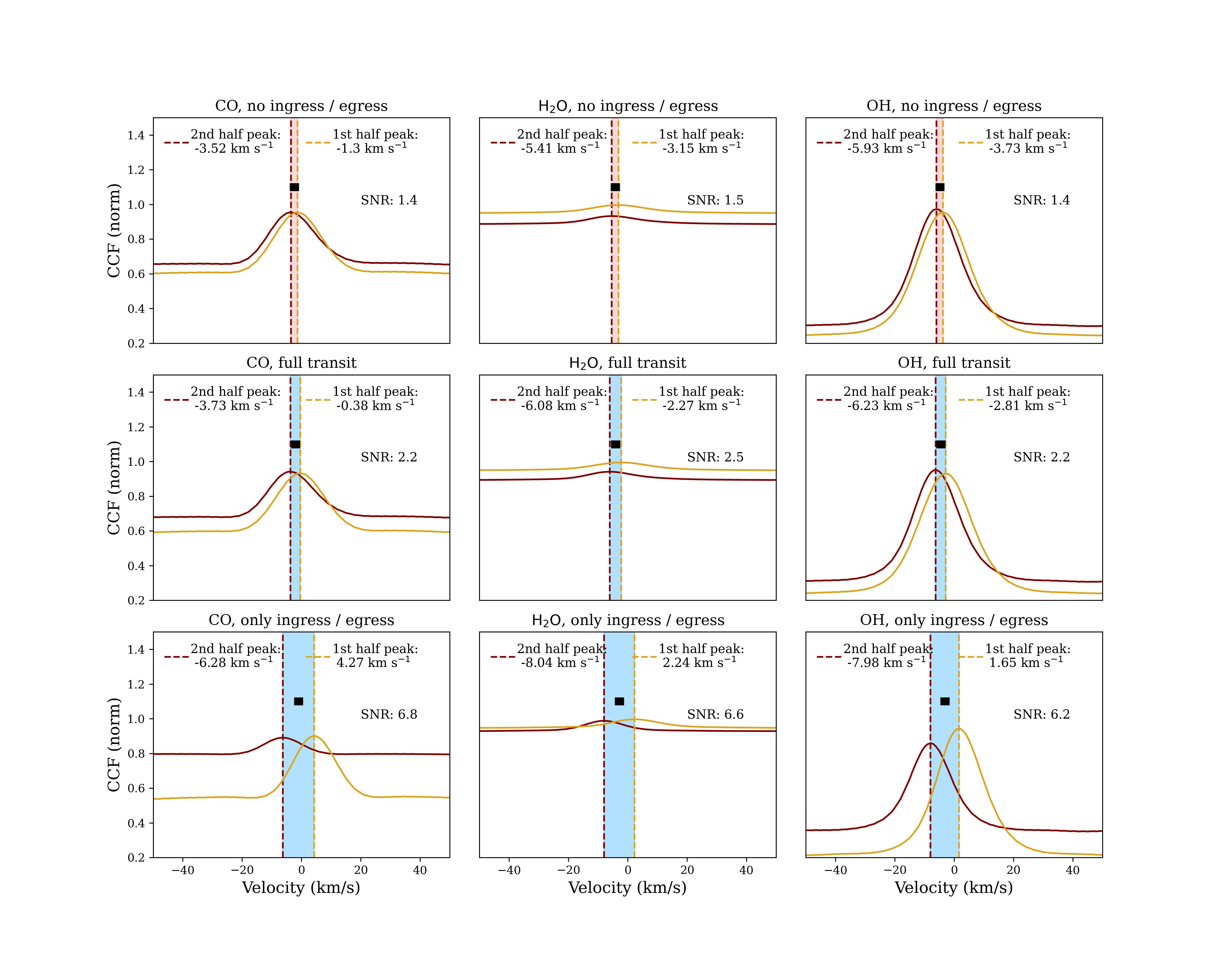}
    \caption{CCFs of individual species averaged over two phase bins. Each column corresponds to different species (OH, CO, $\rm H_2O$), and each row corresponds to different bin selection: without including ingress and egress, including the full transit, and only including ingress and egress. Central bars between the CCFs are colored blue if the difference between the CCFs is greater than optimal Doppler shift errors \citep[1.55 km/s, in black;][]{kesseli2022atomic}; otherwise, they are colored red. In our models, CO only displays detectable CCF differences when only including ingress and egress. The SNR in each plot refers to the difference between the two phase bins' CCF peaks relative to the optimal Doppler shift errors.}
    \label{fig:two_bin_nir}
\end{figure*}

 Figure~\ref{fig:two_bin_nir} shows the results of this experiment. As the error for each phase bin, we take the average error of phase bins from \cite{kesseli2022atomic} (1.55~km\,s$^{-1}$). We define the two phase bins as inconsistent if the peak of their respective CCFs are inconsistent at 2$\sigma$.

We find that excluding ingress and egress phases can strongly reduce the difference in derived Doppler shift between phase bins. Furthermore, we find that, as expected from \cite{kempton2012constraining}, differences between bins are maximized when just considering the ingress and egress phases.\footnote{We would expect that binning with fewer spectra (just including the ingress phases) would increase the associated error on Doppler shift at each bin. However, the point of this exercise is to illustrate the magnitude of ingress/egress Doppler shift discrepancy; observational strategies such as stacking multiple transits could reduce errors in practice and make these differences discernible.}

While higher-order drivers of asymmetry are clearly not detectable with phase bins \citep[e.g,. at what longitude condensation may begin to play a role;][]{wardenier2021decomposing}, certain drivers of asymmetry are accessible with this method. For example, ignoring for now the exact details of error budgets, all species in Figure~\ref{fig:two_bin_nir} clearly blueshift over the course of transit. This provides potential evidence for, among other things, a spatially varying wind field, condensation, optically thick clouds, or a scale height effect. Furthermore, per the results of Section~\ref{sec:co_baseline} the detection of CO's blueshifting indicates that something besides equilibrium chemistry is driving at least some of the asymmetry in the atmosphere. These underlying models are cloud-free, so these results imply sensitivity to, e.g., the scale height effect.

\subsubsection{Comparison to \cite{kesseli2022atomic}}

\begin{figure*}
    \centering
    \includegraphics[scale=0.7]{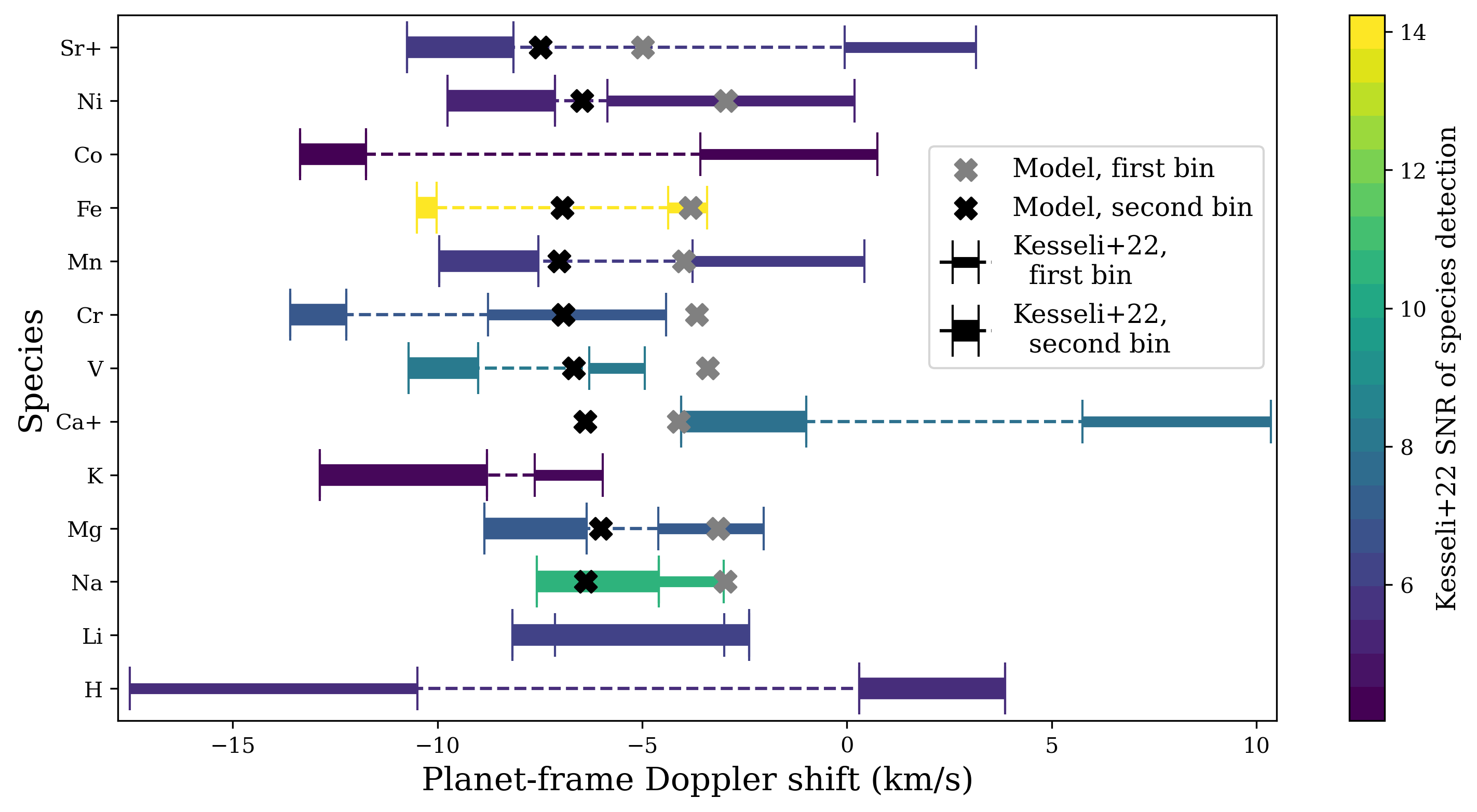}
    \caption{The net Doppler shifts of \cite{kesseli2022atomic} (error bars) as compared to this work's models (crosses). The first phase bin is drawn thinner than the second phase bin; observed phase bins are connected by a dotted line for visibility's sake. The species are ordered and colored by total observed detection SNR. Rows without crosses correspond to species that we could not recover via cross-correlation in our models. Our models are able to explain some species (e.g., Na), fail to explain others (e.g., Cr) and fail to detect yet others (e.g,. K).}
    \label{fig:aurora_comparison}
\end{figure*}

With our models calculated, we can now explore the ability of phase-resolved spectra to confront toy models by comparing the models to observations. A prime observational work that made use of phase binning is \cite{kesseli2022atomic}; there, the authors search for asymmetries in two phase bins for a wide variety of species, motivated by the strength of those species' opacity functions in the data's wavelength range.

To consider a toy model: based on previous studies \citep{ehrenreich2020nightside, tabernero2021espresso, savel2022no}, it appears that Ca~II does not follow the Fe-like Doppler shift trend first observed by \cite{ehrenreich2020nightside}. Rather, it appears that Ca~II, with its strong opacity and resultant deep lines, may be probing a non-hydrostatic region of the atmosphere \citep{casasayas2021carmenes, deibert2021detection, tabernero2021espresso}. This region of the atmosphere cannot be captured by the models of this work and \cite{savel2022no}. 


Without a model of atmospheric escape, it seems difficult to elevate the above picture beyond ``toy model'' status. However, by phase-resolving multiple species, a clearer picture can emerge.

For our comparison with \cite{kesseli2022atomic}, we use the same line lists as in that study: the National Institute of Standards and Technology \citep[NIST;][]{kramida2019nist} line lists. It is crucial to use the same line lists for comparisons of HRCCS studies---different line list databases can contain vastly discrepant numbers of line transition, which greatly affects the resultant opacity function (see, for instance, Figure 11 of \citealt{grimm2021helios}).

The results of our comparison with the species detected in \cite{kesseli2022atomic} are shown in Figure~\ref{fig:aurora_comparison}. As in \cite{savel2022no}, these baseline models---no clouds, no condensation, no orbital eccentricity---cannot fully explain the Doppler shifts of Fe observed in WASP-76b. However, the comparison across multiple different species provides further constraints. Figure~\ref{fig:aurora_comparison} shows that Fe, V, Cr, Ca~II, and Sr~II are strongly discrepant from our models for at least one half of transit, whereas Na, Mg, Mn, and Ni are reasonably well described by our models for both the first and second half of transit. Furthermore, Fe, V, and Cr all have stronger blueshifts in the second phase bin than in our models. The similar level of disagreement between Fe, V, and Cr implies that they share a common driver of asymmetry. This result in turn implies that whatever driver affects them affects the regions in which these species form similarly --- be it clouds, condensation, etc.

\addressresponse{To bridge the toy models presented in Section~\ref{sec:asymmetry_chemistry} to our \cite{kesseli2022atomic} comparison, we compute a set of high-resolution spectra exactly as above, but with the same altitude grid at all latitudes and longitudes in an effort to effectively turn off the scale height effect while maintaining chemical limb inhomogeneities. Post-processing this (self-inconsistent) model yields less than half the Doppler shift asymmetry as compared to our self-consistent models. This experiment confirms the intuition that the scale height effect is a first-order asymmetry effect.}

\begin{figure*}
    \centering
    \includegraphics[scale=0.4]{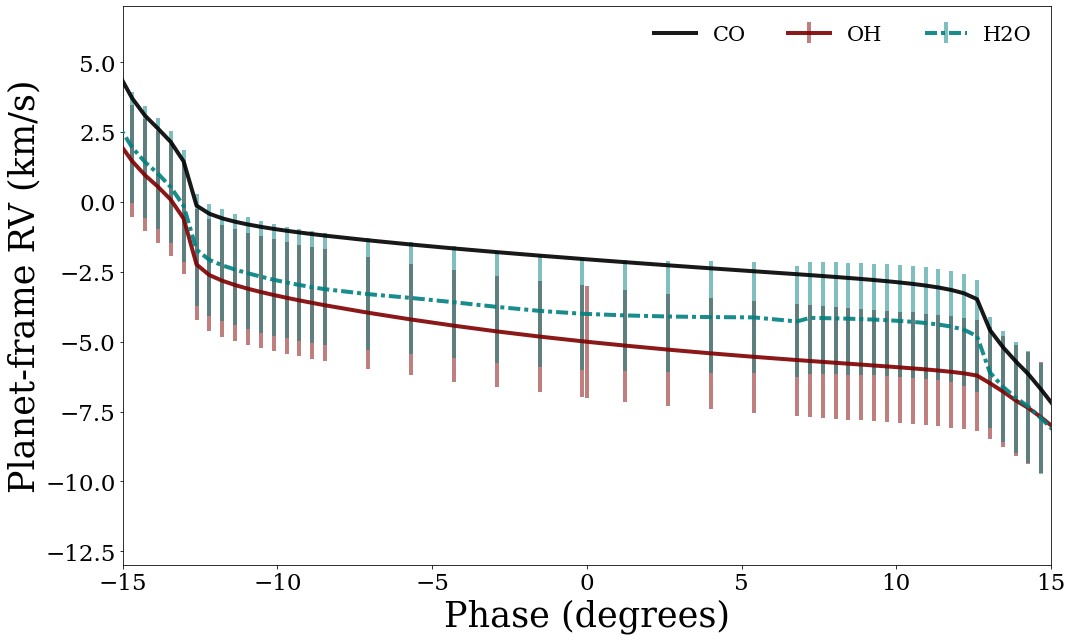}
    \caption{Modeled phase-resolved Doppler shifts for select NIR-absorbing species, with representative error bars \citep{ehrenreich2020nightside} drawn on. We find that OH and $\rm H_2O$ have distinct Doppler signatures from CO; however, OH and $\rm H_2O$ have Doppler shifts that are indistinguishable from one another with current best-case error bars \citep[e.g.,][]{ehrenreich2020nightside}. Considering CO as a ``baseline species'' here allows one to better understand how $\rm H_2O$ and OH may change through the atmosphere.}
    \label{fig:nir_doppler}
\end{figure*}

Finally, we consider the Ca~II toy model previously described. Certain lightweight and/or ionized species may be entrained in an outflow, as indicated by some previous observations \citep[e.g.,][]{tabernero2021espresso} of very deep absorption lines in transmission that must extend very high up in altitude. The differential behavior of the Ca~II and Sr~II Doppler shifts lends more credence to this hypothesis.

In sum, by taking advantage of phase-binned spectra, it is possible to better identify drivers of HRCCS asymmetry. Additionally, our predictions in Figure~\ref{fig:aurora_comparison} indicate that most species should have roughly the same Doppler shift patterns. In stark contrast, observations reveal much larger variations in velocity across different species. While some interpretation may be due to spurious detections, physics that is not included in our model (e.g., outflows, condensation) may be playing a driving role.

\subsection{Full phase-resolved spectra}
Currently, the most information-rich diagnostic available to probe asymmetry in HRCCS is phase-resolved cross-correlation functions \citep[e.g.,][]{ehrenreich2020nightside,borsa2021atmospheric}---that is, net Doppler shifts associated with the absorption spectrum evaluated over multiple points in transit. With these data, one should be able to directly constrain longitudinally dependent drivers of asymmetry, providing the best chance of disentangling the physical mechanisms outlined in Section \ref{sec:drivers}. But how far can we push these data?

\subsubsection{Example: probing physics in the NIR}\label{sec:phase_resolved_nir}

To explore this question, we take as an example a three-species (OH, $\rm H_2O$, and CO) near-infrared (NIR) dataset over a CARMENES-like waveband as in Section~\ref{sec:phase bins}. Figure~\ref{fig:nir_doppler} shows the Doppler shifts of these species as a function of phase, produced for single species at a time as in Section \ref{sec:phase bins}, but without any averaging. 

Without considering any data, a compelling toy model would be as follows: $\rm H_2O$ is thermally dissociated on the hotter, approaching limb, so it preferentially exists on the receding limb. OH is a product of $\rm H_2O$ photodissociation, so it forms preferentially on the approaching limb. CO is constant everywhere; therefore, CO should not experience much of a trend in Doppler shift, OH should be more blueshifted than CO, and $\rm H_2O$ should be more redshifted than CO.

We shall see, however, that additional, complicating physics is revealed by fully phase-resolved spectra. For our models, the relevant underlying physics is as follows:

\begin{figure*}
    \gridline{\fig{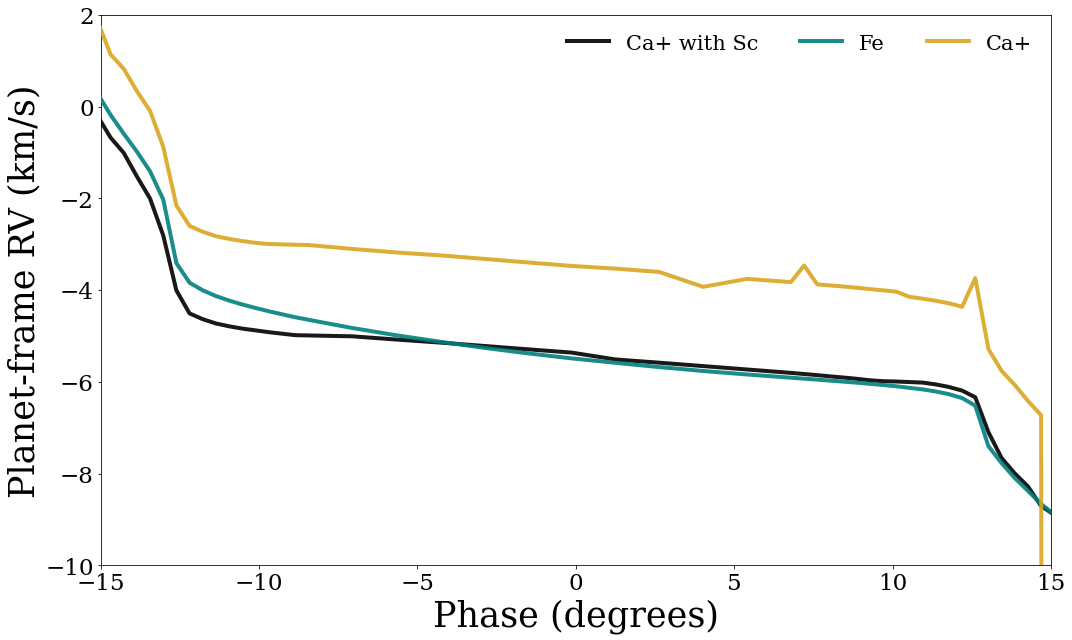}{0.5\textwidth}{(a)}
          \fig{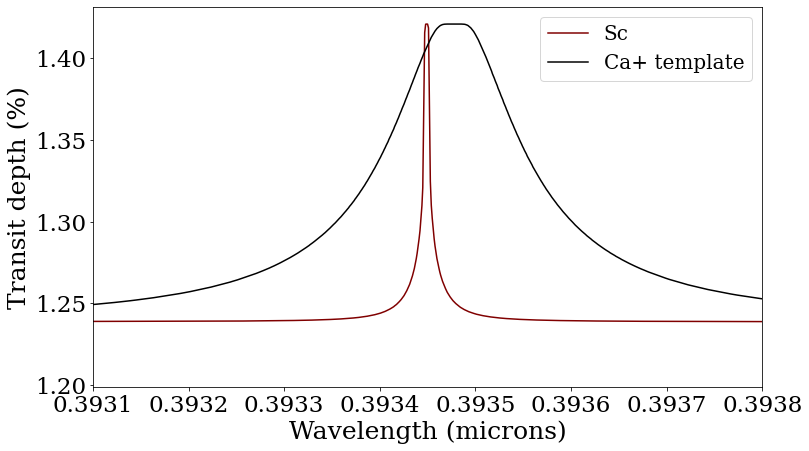}{0.5\textwidth}{(b)}
          }
\caption{Results of an investigation into anomalous Ca II blueshift between different model runs. In panel (a), it can be seen that forward models that include absorption due to Sc opacity yield a larger Ca II blueshift than models that lack Sc (Fe Doppler shift is included for comparison). Panel (b) illustrates the cause of this anomalous blueshift: a Sc line overlapping one line in the optical Ca II doublet. These results imply that overlapping line profiles can subtly contaminate calculated Doppler shifts.
\label{fig:ca+_sc}}
\end{figure*}
\begin{enumerate}
    \item \textbf{Altitude-dependent winds}: $\rm H_2O$ lines are more strongly blueshifted than CO lines at all phases because the $\rm H_2O$ line cores over the wavelength range of the CARMENES bandpass more predominantly form at higher altitudes. At high altitudes, the atmospheric flow switches from dominantly rotational (via an eastward equatorial jet) to dominantly divergent (via day--night winds) \citep{hammond2021rotational}. This result is the opposite of what would be expected from the above-described toy model, revealing the shortcomings of simple models and how they can sometimes mislead us.
    \item \textbf{Equilibrium chemistry}: $\rm H_2O$ and CO are less blueshifted than OH because OH preferentially forms on the approaching, blueshifted limb of the planet. OH being more blueshifted than the other molecules is in agreement with the predictions of the toy model.
    \item \textbf{Equilibrium chemistry}: The relationship between $\rm H_2O$ and OH changes as a function of phase because the ratio $\rm OH / H_2O$ increases a function of temperature, and hotter regions of the planet rotate into view over transit. This finding is also qualitatively in agreement with the toy model. However, per Figure~\ref{fig:nir_doppler}, this effect is unfortunately not likely to be observable given the error bars in current data sets.
\end{enumerate}



Now the question remains: Can we observe in real data the trends matching these model explanations? As a simple experiment, we can apply error bars representative of the best observing nights on the best instrument with the most observable chemical species (roughly 2 km/s, as drawn as vertical error bars in Figure~\ref{fig:aurora_comparison}; \citealt{ehrenreich2020nightside}) and determine whether these trends are still detectable. With our errorbars now applied to our simulated data, only the first explanation---that $\rm H_2O$ forms at higher altitudes than CO---can fully be addressed, assuming that Doppler shifts for both species can be obtained. The second explanation can only be partially addressed---we can still determine that CO is less blueshifted than OH.


\subsubsection{Warning: blending of Doppler shifts}
The disentangling of physics in Section~\ref{sec:phase_resolved_nir} rests on a fundamental assumption: that each cross-correlation template directly tracks only a single species. Indeed, one of the promises of HRCCS is the ability to uniquely constrain individual species' abundance; with individual line profiles resolved, different species should be readily identifiable from one another in cross-correlation space \citep[e.g.,][]{brogi2019retrieving}. Furthermore, our noiseless models should be even less susceptible to degeneracies between different species' spectral manifestations.

Panel (a) of Figure~\ref{fig:ca+_sc} seems to contradict the notion of complete line profile independence across species. For models run in \cite{savel2022no}, Sc was excluded. Motivated by the search for atoms in \cite{kesseli2022atomic}, however, we included Sc in this work's models. Surprisingly, we found a subsequent significant difference in the Doppler shifts recovered from our cross-correlation analysis in our Sc-inclusive models. 

Panel (b) of Figure~\ref{fig:ca+_sc} reveals the source of the discrepancy. In the optical, Ca II opacity is dominated by a doublet; one of the lines in this doublet partially overlaps with a strong, narrow Sc line. When both species combined in a forward model, the Sc line produces absorption just blueward of this Ca II line's core; hence, the cross-correlation of the Ca II template yields a spurious blueshift. There did exist other modeling differences between the two spectra (e.g., the \cite{savel2022no} models included TiO and VO), but none of these differences strongly impacted the Doppler shift of Ca~II.

Because Ca II in the optical has only two strong lines, it is particularly susceptible to this type of error. All it takes is one slight overlap with another species near a Ca II doublet core, and the Ca II Doppler signal can be significantly biased. Species with forests of lines (e.g., Fe in the optical) should hence be more robust to chance overlaps with other species' lines.

To guard against this error for species with few lines, we recommend cross-correlating templates against one another to get a first-order sense for the extent of species overlap in Doppler space. Furthermore, we recommend performing these analyses on HRCCS with combined species models, as opposed to single-species models. \addressresponse{This approach could involve a retrieval framework \citep{brogi2019retrieving, gandhi2019hydra,gibson2020detection}, which couples a statistical sampler to an atmospheric forward model to determine the exoplanet spectrum that best fits the data, inclusive of multiple chemical species at once.}


\section{Conclusion} \label{sec:conclusion}
The past few years have yielded asymmetric Doppler signals from exoplanet atmospheres as a function of phase. Compelling ``toy models'' notwithstanding, a number of physical processes can drive these asymmetries, and it can be difficult to uniquely constrain the cause of an asymmetry.

In this study, we determine that if an asymmetry is observed:

\begin{enumerate}
   \item It may be due to a scale height difference across the atmosphere, not a chemistry difference across the atmosphere. Comparing a signal of a species in HRCCS to a baseline species that is guaranteed to be chemically stable over the atmosphere can better motivate whether the asymmetry could be due to chemistry. CO is an excellent baseline species for ultra-hot Jupiters, as it is stable over these planets' expected temperature--pressure space, has many spectral lines in the near-infrared accessible to ground-based spectrographs, and has been detected in numerous studies.
  \item The asymmetry can be highly informative even if it is binned in phase, especially if multiple species are considered. For instance, much larger Doppler shifts (both blue and red) of certain species relative to the predictions of hydrostatic GCMs can be used as evidence for outflowing material.
   \item The asymmetry may be boosted by including (and perhaps only considering) ingress and egress phases. Ingress and egress spectra are the the gold standard for asymmetric signals so long as the signal to noise is high enough.
   \item The asymmetry may be influenced by line confusion between species, even at high resolution. Species with very few lines (e.g., a single doublet) in the observed waveband are especially susceptible to contamination by other species in cross-correlation analysis, and they should be carefully checked against theoretical models for possible contaminating opacity sources.
   \item If all species exhibit a similar asymmetry---especially if they all become less blueshifted over the course of transit---the asymmetry may be due to a large-scale effect, such as clouds blanketing the cooler limb.
   \item Per our comparison of near-infrared absorbers in the CARMENES waveband, the toy model predictions of the $\rm H_2O$ Doppler shift relative to CO was inaccurate, as it did not include information about the vertical coordinate. With $\rm H_2O$ lines on average probing higher in the atmosphere than CO in this waveband, they probed a different part of the flow, departing from expectations of the toy model.
\end{enumerate}

By aiming to systematically understand even just a few drivers of asymmetry, this work has made it clear that HRCCS---already arguably abstract given its general inability to produce visible planetary spectra---has yet more nuance to uncover. As data quality continues to increase, it will become increasingly necessary to understand the relationships between higher-order physical effects. 

\begin{acknowledgments}

A.B.S., E.M.-R.K., and E.R. acknowledge funding from the Heising-Simons Foundation.

We thank Michael Zhang for a thoughtful conversation on the cross-correlation signature of HCN. We also thank Anusha Pai Asnodkar for a robust discussion of degeneracies in HRCCS tests. Finally, we thank Serena Cronin for providing useful insight into applications of CO detections in extragalactic astronomy.

The authors acknowledge the University of Maryland supercomputing resources (\url{http://hpcc.umd.edu}) made available for conducting the research reported in this paper.

This research has made use of NASA’s Astrophysics Data System Bibliographic Services.
\end{acknowledgments}

\software{\texttt{astropy} (\citealt{astropy:2018}), \texttt{batman} \citep{kreidberg2015batman}, \texttt{FastChem} \citep{stock2018fastchem}, \texttt{IPython} (\citealt{perez2007ipython}), \texttt{HELIOS-K} \citep{grimm2021helios}, \texttt{HELIOS} \citep{malik2017helios}, \texttt{Matplotlib} (\citealt{hunter2007matplotlib}), \texttt{NumPy} (\citealt{2020NumPy-Array}), \texttt{Numba} \citep{lam2015numba},
\texttt{pandas} (\citealt{mckinney2010data}),  
\texttt{SciPy} (\citealt{virtanen2020scipy}), \texttt{tqdm} (\citealt{da2019tqdm})}



\bibliography{main}{}
\bibliographystyle{aasjournal}



\end{document}